\documentclass[reqno]{amsart}
\usepackage{amsmath,amssymb}
\usepackage{graphics,epsfig,color}
\usepackage[mathscr]{euscript}
\theoremstyle{plain}

\theoremstyle{definition}

\theoremstyle{remark}

\numberwithin{equation}{section}
\numberwithin{theorem}{section}
\renewcommand{\epsilon}{\varepsilon}
\renewcommand{\tilde}{\widetilde}
\renewcommand{\hat}{\widehat}
\definecolor{light}{gray}{.9}

\title[Totally asymmetric limit]{Totally asymmetric limit for models of heat conduction }

\author[L.\ De Carlo]{Leonardo De Carlo}
\address{Leonardo De Carlo \hfill\break \indent
  GSSI
  \hfill\break\indent
  Viale Francesco Crispi 7,   67100 Coppito, L'Aquila, Italy}
\email{neoleodeo@gmail.com}

\author[D.\ Gabrielli]{Davide Gabrielli}
\address{Davide Gabrielli \hfill\break \indent
  DISIM, Universit\`a dell'Aquila
  \hfill\break\indent
  Via Vetoio,   67100 Coppito, L'Aquila, Italy}
\email{gabriell@univaq.it}

\begin{document}

\begin{abstract}
We consider one dimensional weakly asymmetric boundary driven models of heat conduction. In the cases of a constant diffusion coefficient and of a quadratic mobility we compute the quasi-potential that is a non local functional obtained by the solution of a variational problem. This is done using the dynamic variational approach of the macroscopic fluctuation theory \cite{MFT}. The case of a concave mobility corresponds essentially to the exclusion model that has been discussed in \cite{Lag,CPAM,BGLa,ED}. We consider here the convex case that includes for example the Kipnis-Marchioro-Presutti (KMP) model and its dual (KMPd) \cite{KMP}.
This extends to the weakly asymmetric regime the computations in \cite{BGL}.
We consider then, both microscopically and macroscopically, the limit of large external fields. Microscopically we discuss some possible totally asymmetric limits of the KMP model. In one case the totally asymmetric dynamics has a product invariant measure. Another possible limit dynamics has instead a non trivial invariant measure for which we give a duality representation. Macroscopically we show that the quasi-potentials of KMP and KMPd, that for any fixed external field are non local, become local in the limit. Moreover the dependence on one of the external reservoirs disappears. For models having strictly positive quadratic mobilities we obtain instead in the limit a non local functional having a structure similar to the one of the boundary driven asymmetric exclusion process.

\bigskip

\noindent {\em Keywords}: Non equilibrium statistical mechanics, large deviations, stochastic lattice gases.

\smallskip

\noindent{\em AMS 2010 Subject Classification}:
60F10, 60K35, 82C05  %bho

\end{abstract}

\maketitle

\section{Introduction}

Understanding the structure of stationary non equilibrium states is a major issue in non equilibrium statistical mechanics. In recent years some one dimensional boundary driven stochastic lattice models have been exactly solved. A stationary non equilibrium state can be described microscopically exhibiting the invariant measure of the model or macroscopically describing the structure of the fluctuations. This second approach is usually based on large deviations theory and the corresponding rate functional has some thermodynamic interpretations \cite{MFT,D}. Apart some special cases, the rate functional is non local or correspondingly the invariant measure has long range correlations.

Stochastic lattice gases for which the hydrodynamic scaling limit has been derived are essentially of two types. Driven diffusive models that have a non trivial diffusive scaling limit typically given by a non linear diffusive equation \cite{KL99,Spohn} and asymmetric models whose natural scaling is the Eulerian one and have as a scaling limit a first order hyperbolic conservation law \cite{Bah1,KL99,Spohn}. There are examples of solvable one dimensional stationary non equilibrium states in both classes of models.

In the case of diffusive systems, the first computation of large deviations rate functionals was obtained for the exclusion model starting from an exact representation of the invariant measure \cite{D,DLSA,DLSB}. The same result was then obtained with the dynamic variational approach of the macroscopic fluctuation theory \cite{BDGJL1,BDGJL2,MFT}. This macroscopic approach was then generalized to a wider class of models characterized by a constant diffusion matrix and a quadratic mobility \cite{MFT,BGL}.

For the weakly asymmetric exclusion the computation of the rate functional has been done in \cite{ED} starting from an exact representation of the invariant measure and then in \cite{Lag,CPAM,BGLa} using the macroscopic fluctuation theory. A weakly asymmetric model is a microscopic model having a behavior whose asymmetry is suitably going to zero when the underlying lattice size is going to zero. The weak asymmetry
is encoded in the limiting behavior by an external field $E$.

In the case of asymmetric models the situation is somehow less developed. The first computation of a large deviation rate functional for a stationary non equilibrium state has been obtained for the boundary driven asymmetric exclusion process using an exact representation of the invariant measure \cite{DLS0,DLS}. The rate functional is not local and has a different representation depending if the asymmetry is driving in the same or in the opposite direction with respect to the boundary sources. In this case the dynamic variational approach is more difficult to be applied since it is more difficult to identify the basic ingredient that is the dynamic large deviations functional. In the case of the exclusion process the dynamic rate functional has been obtained with periodic boundary conditions by L. Jensen in his Ph.D. thesis (see \cite{V}) but the proof is not complete.
The case with boundary sources has been discussed in \cite{BD}. Once the dynamic rate functional is identified then it is possible to define and in some cases to compute the corresponding quasi-potential \cite{Bah2,MFT}. In particular in \cite{Bah2} the functionals in \cite{DLS0,DLS} for the exclusion process have been obtained using the dynamic variational approach. In \cite{Bah2} the quasi-potential has been computed also for other conservation laws with a convex mobility.

Other examples of exact computations of large deviations rate functionals for stationary non equilibrium states are for example \cite{Ber} for a two component diffusive system and \cite{D} for the totally asymmetric exclusion process with particles of different classes on a ring.

A different way of obtaining the functionals in \cite{DLS0,DLS} for the asymmetric exclusion process is to compute the functional for the weakly asymmetric case and then to consider the limit for large values of the field $E$. This has been done in \cite{Lag,CPAM,BGLa}.
With this approach it has been also proved the existence of Lagrangian phase transitions for finite but large external fields \cite{Lag,CPAM}. In this paper we use exactly this approach generalizing it to the case of models having constant diffusion matrix and convex quadratic mobility.

We consider also the problem from a microscopic perspective. In the case of the symmetric simple exclusion, switching on an external field we obtain an asymmetric exclusion model. In the limit of a very large external field the mass can flow only in one direction and the only natural limiting process is the totally asymmetric exclusion process. In the case of a model like the KMP the situation is more complex. In the totally asymmetric limit the mass can flow only in one direction but in this case there are several possible natural  dynamics satisfying this constraint. Since their  scaling limit in the weakly asymmetric regime is determined by the behavior for small values of the external field, all of them will have the same weakly asymmetric scaling limit. We discuss two possible totally asymmetric versions of the KMP model. In one case the invariant measure for the boundary driven case is given by a product measure of exponentials whose parameter is determined by the chemical potential of just one external source. For the other dynamics the invariant measure in the boundary driven case has apparently not a simple structure and we give a representation in terms of convex combinations of products of Gamma distributions. This is done using a duality representation of the process constructed using convex analytic arguments. The same could be done also for the dual model of the KMP (that we shortly call KMPd).

Macroscopically we start computing the large deviations rate functional for the empirical measure when the particles are distributed according to the invariant measure of weakly asymmetric models of heat conduction. Instead of discussing the general case we concentrate on three prototypal models corresponding respectively to mobilities having two coinciding roots, two distinct roots and no roots. For the first case we choose the special form of the mobility $\sigma(\rho)=\rho^2$ that microscopically corresponds to the KMP model. For the second case we choose the special form of the mobility $\sigma(\rho)=\rho(1+\rho)$ that microscopically corresponds to the dual of the KMP model (KMPd). For the third case we choose the special form of the mobility $\sigma(\rho)=1+\rho^2$. We do not have an explicit example of a microscopic model corresponding to this form of the mobility. We conventionally call this model KMPx. In this case the density can assume also negative values. For all these models, in the one dimensional case, we can compute the quasi-potential using the dynamic variational approach of the macroscopic fluctuation theory \cite{MFT}. The corresponding rate functionals are generically not local and depends on the value of the constant external field $E$ generating the asymmetry.  For each choice of the external sources there is however a special value $E^*$ of the field for which the system is an equilibrium one so that there is no current in the stationary state and the large deviations functional is local. The rate functional is computed solving the corresponding infinite dimensional Hamilton Jacobi equation. As in the case of the simple exclusion model \cite{Lag,CPAM,BGLa} depending if the value of the external field is above or below the threshold $E^*$ we have that the rate functional is represented as a supremum or an infimum of an auxiliary functional.

Once computed the quasi-potentials depending on the parameter $E$ we study the corresponding
asymptotic behavior when $E\to\pm\infty$. In the case of the exclusion model this limits allow to recover the large deviations rate functionals of the totally asymmetric exclusion process \cite{Lag,CPAM,BGLa}. In the case of KMP and KMPd we have that the functionals, that for each finite value of $E$ are non local, become local in the limit. Moreover the dependence on the left source disappears in the limit $E\to-\infty$ while the dependence on the right one disappears in the limit $E\to +\infty$. This is quite natural having in mind the underlying microscopic stochastic dynamics. More precisely the limit functionals correspond to the large deviations rate functionals for empirical measures when the variables are distributed according to product measures. In the case of KMP the single marginal is exponential and in the KMPd case is instead geometric. We recover in this way the large deviations rate functionals for one of the two versions of the totally asymmetric KMP dynamics. The model KMPx has instead a very different behavior. The limiting functional is again non local and has a structure very similar to the one of the exclusion. In this case the possibility of Lagrangian phase transitions appears for large values of the field acting in the same direction with respect to the external reservoirs.

\smallskip
The structure of the paper is the following

In Section 2 we describe microscopically the symmetric versions of the models of heat conduction that
we are going to study, describe the weakly asymmetric versions, study the invariant measures and describe the instantaneous current that is the basic object to understand the connection between the microscopic and the macroscopic description.

In Section 3 we introduce two possible totally asymmetric versions of the KMP dynamics and study the corresponding invariant measures in the boundary driven case. For one model the invariant measure is of product type while for the other one this is not the case and we give a duality representation using convex analytic arguments.

In Section 4 we give a short overview of the scaling limit for particle systems that is the main bridges between the microscopic and macroscopic descriptions. We discuss the transport coefficients and introduce macroscopically the three prototypal models we are going to study.

In Section 5 we study the stationary solutions of the hydrodynamic equations and the corresponding associated currents. In particular we discuss the asymptotic behaviors for large fields $E$.

In Section 6 we outline the structure of dynamic large deviations for our class of models and show the relation between the quasi-potential and the large deviations for the invariant measure.

In Section 7 we compute the quasi-potential for  boundary driven one dimensional weakly asymmetric models.

In Section 8 we compute the limit of the quasi-potentials for $E\to \pm \infty$.

\section{Models of heat conduction}

In this section we introduce the symmetric versions of the stochastic microscopic dynamics that we are going to study. We introduce also the weakly asymmetric perturbations, discuss reversibility and introduce the instantaneous current.

\subsection{Symmetric models}
We start describing the symmetric version of the Kipnis-Marchioro-Presutti (KMP) model \cite{KMP}.
This is a generalized stochastic lattice gas on which energies associated to oscillators located at the vertices of a lattice randomly evolve. More precisely let $\Lambda\subseteq \mathbb R^d$ be a bounded domain and let $\Lambda_N:=\frac 1N \mathbb Z^d \cap \Lambda$ be its discretization with a lattice of mesh $\frac 1N$. The oscillators are located at the vertices of the lattice and we denote by $\xi=\left\{\xi(x)\right\}_{x\in \Lambda_N}$ the energy configuration of the system. The value $\xi(x)\in \mathbb R^+$ is the energy associated to the oscillator at site $x$. The interpretation of the configuration $\xi$ as a configuration of energy follows by the original definition of the model \cite{KMP}. Since we are discussing also generalizations and different models we prefer to switch the interpretation to a configuration of mass.  We will mainly consider the one dimensional case for which $\Lambda$ is an interval and $\Lambda_N$ is a linear lattice. We call $x\in \Lambda_N$ an internal vertex if all its nearest neighbors $y\in \frac 1N\mathbb Z^d$ belong also to $\Lambda_N$. A vertex $x\in \Lambda_N$ that is not internal is instead a boundary vertex. We denote by $\partial \Lambda_N$ the set of boundary vertices. The stochastic evolution is encoded in the generator that is of the type
\begin{equation}\label{genKMP}
\mathcal L_N f =\sum_{x\sim y}L_{x,y}f +\sum_{x\in \partial \Lambda_N}L_x f\,.
\end{equation}
The first sum in \eqref{genKMP} is a sum over unordered nearest neighbor sites of $\Lambda_N$. The first term in \eqref{genKMP}
is the bulk contribution to the stochastic evolution while the second term is the boundary part of the dynamics that modelizes the interaction of the system with external reservoirs.

We introduce the model using a slightly different language from the classic one. This approach simplifies notation and is suitable for generalizations. Let $\varepsilon^x=\left\{\varepsilon^x(y)\right\}_{y\in \Lambda_N}$ be the configuration of mass with all the sites different from $x$ empty and having unitary mass at site $x$. This means that $\varepsilon^x(y)=\delta_{x,y}$ where $\delta$ is the Kronecker symbol. The bulk contribution to the stochastic dynamics is defined by
\begin{equation}\label{bulkxy}
L_{x,y}f(\xi):=\int_{-\xi(y)}^{\xi(x)}\frac{dj}{\xi(x)+\xi(y)}\big[f(\xi-j\left(\varepsilon^x-\varepsilon^y\right))-f(\xi)\big]\,.
\end{equation}
It is immediate to check that definition \eqref{bulkxy} is symmetric in $x,y$ so that we can consider without ambiguity a sum over unordered pairs in \eqref{genKMP}. Formula \eqref{bulkxy} define the model as a random current model. The intuition behind the formula is the following. On each bond of the system there is a random flow of mass that happens accordingly to a random exponential clock with rate $1$. When this random clock rings there is a flow between the two endpoint sites that is uniformly distributed among all the possible currents that keep the masses at the two sites positive. The new configuration $\xi-j\left(\varepsilon^x-\varepsilon^y\right)$ is the starting configuration minus the divergence of a current on the lattice different from zero on the single edge $(x,y)$ where it assumes the value $j$. The choice of a uniformly random current in \eqref{bulkxy} corresponds to the usual KMP dynamics.

The boundary part of the generator can be defined in several ways. Let us fix a possible definition that is good for symmetric and weakly asymmetric models. Consider the left boundary of a one dimensional system on $\Lambda=(0,1)$. The system is in contact with an external reservoir with chemical potential $\lambda<0$. The effect of the interaction with the source is that at rate one the value of the variable $\xi(1/N)$ is substituted by a random value exponentially distributed with parameter $-\lambda$. When the value $\xi(1/N)$ is substituted by the value $z$ we imagine that there is a current $j=z-\xi(1/N)$ across the edge
$\left(0,1/N\right)$. With a change of variables we can then write the boundary term of the generator with a random current representation as
\begin{eqnarray}\label{bkmps}
& & L_{1/N}f(\xi)=\int_0^{+\infty}|\lambda| e^{\lambda z}\left[f(\xi+\varepsilon^{1/N}(z-\xi(1/N)))-f(\xi)\right]\,dz= \nonumber \\
& &  \int_{-\xi(1/N)}^{+\infty}|\lambda| e^{\lambda (\xi(1/N)+j)}\left[f(\xi+\varepsilon^{1/N}j)-f(\xi)\right]\,dj\,.
\label{arpaceltica}
\end{eqnarray}
The dynamics \eqref{bulkxy} can be generalized substituting the uniform distribution on $[-\xi(y),\xi(x)]$ with
a different probability measure (or just positive measure) $\Gamma_{x,y}^\xi(dj)$ on the same interval.
A natural choice is the discrete uniform distribution on the integer points in $[-\xi(y),\xi(x)]$. This means that if $\xi$ is a configuration of mass assuming only integer values then
\begin{equation}\label{kmpdu}
\Gamma_{x,y}^\xi(dj)=\frac{1}{\xi(x)+\xi(y)+1}\sum_{i\in [-\xi(y),\xi(x)]}\delta_i(dj)
\end{equation}
where $\delta_i(dj)$ is the delta measure concentrated at $i$ and the sum is over the integer values belonging to the interval.
If the initial configuration is such that the values of the variables $\xi$ are all integers then this fact is preserved by the dynamics and we obtain a model that can be interpreted as a model of evolving particles. This is exactly the dual model of KMP \cite{KMP}. We call the stochastic dynamics associated to the choice \eqref{kmpdu} KMPd where the last letter means \emph{dual}. The boundary dynamics can be fixed similarly to \eqref{bkmps}. In this case it is natural to substitute the exponential distribution by a geometric one.

Another interesting model could be related to Gaussian distributions. In this case the interpretation in terms of mass is missing since the variables can assume also negative values. The bulk dynamics is defined by a distribution of current having support on all the real line and defined by
\begin{equation}\label{currga}
\Gamma_{x,y}^\xi(dj)=\frac{1}{\sqrt{2\pi \gamma^2}}e^{-\frac{\left(j-\frac{(\xi(x)-\xi(y))}{2}\right)^2}{2\gamma^2}}dj\,.
\end{equation}
In general we use the same notation both for a  measure and the corresponding density.
Note that all these models share the symmetry $\Gamma^\xi_{x,y}(j)=\Gamma^\xi_{y,x}(-j)$ and for this reason we can write
in \eqref{genKMP} a sum over unordered nearest neighbor sites. Also in this case it is possible to introduce a boundary part of the dynamics.

\subsection{Weakly asymmetric models}

If we switch on a possibly space and time dependent external field, the random distribution of the current is changed. In particular suppose that on the lattice it is defined a discrete vector field $\mathbb F$. This is a collection of numbers $\mathbb F(x,y)$ for any ordered pair of nearest neighbors lattice points satisfying the antisymmetry relationships $\mathbb F(x,y)=-\mathbb F(y,x)$. If the vector field is time dependent these number are time dependent. The motion of the mass is influenced by the presence of the field and we have a model with a random current across each bond distributed according to a measure $\Gamma^{\mathbb F}$ as follows
\begin{equation}\label{bulkxyE}
L_{x,y}^{\mathbb F}f(\xi):=\int\Gamma^{\xi,\mathbb F}_{x,y}(dj)\big[f(\xi-j\left(\varepsilon^x-\varepsilon^y\right))-f(\xi)\big]\,.
\end{equation}
The natural choice of the measure $\Gamma^{\mathbb F}$ is
\begin{equation}\label{gammae}
\Gamma^{\xi,\mathbb F}_{x,y}(dj)=\Gamma^{\xi}_{x,y}(dj)e^{\frac{\mathbb F(x,y)}{2}j}\,.
\end{equation}
The factor $\frac 12$ in the exponent appears just for convenience of notation in the following.
A perturbation of this type is for example the one used in \cite{BGL} to compute dynamic large deviations for the KMP model and  corresponds therefore to the choice
\begin{equation}\label{vecchiapert}
\Gamma^{\xi,\mathbb F}_{x,y}(dj)=\frac{e^{\frac{\mathbb F(x,y)}{2}j}}{\xi(x)+\xi(y)}\chi_{[-\xi(y),\xi(x)]}(j)\,dj\,.
\end{equation}
By the symmetry of the measure $\Gamma$ and the antisymmetry of the discrete vector field $\mathbb F$ we have that $\Gamma^{\xi,\mathbb F}_{x,y}(j)=\Gamma^{\xi,\mathbb F}_{y,x}(-j)$ and we can define the generator considering sums over unordered bonds.

The terminology weakly asymmetric corresponds to the situation when the parameter $N$ is large and the discrete vector field is obtained as a discretization of a smooth vector field on $\Lambda$. In this case since the mesh of the lattice is $\frac 1N$ the values of the discrete vector field are $O\left(\frac 1N\right)$. A natural discretization is as follows.

Let $F: \Lambda\to \mathbb R^d$ be a smooth vector field with components
$$
F(x)=\left(F_1(x),\dots ,F_d(x)\right)\,.
$$.
We associate to $F$ a discrete vector field $\mathbb F$ on the lattice that corresponds to a discretized version of the continuous vector field defined by
\begin{equation}\label{disc-vec}
\mathbb F\left(x,y\right)=\int_{(x,y)} F\left(z\right)\cdot dz \,.
\end{equation}
In \eqref{disc-vec} $(x,y)$ is an oriented edge of the lattice, the integral is a line integral
that corresponds to the work done by the vector field $F$ when a particle moves from $x$ to
$y$. The value of $\mathbb F\left(y,x\right)$, by antisymmetry,  corresponds to minus the value in \eqref{disc-vec}.

\subsection{stationarity}\label{unno}

We have for the symmetric KMP model that the detailed balance condition
\begin{equation}\label{detbal}
\mu_N^\lambda(\xi)\Gamma_{x,y}^\xi(j)=\mu_N^\lambda\big(\xi-j\left(\varepsilon^x-\varepsilon^y\right)\big)
\Gamma_{x,y}^{\xi-j\left(\varepsilon^x-\varepsilon^y\right)}(-j)
\end{equation}
is satisfied when
\begin{equation}\label{protuono}
\mu_N^\lambda(\xi)=\prod_z|\lambda| e^{\lambda\xi(z)}
\end{equation}
is the density of the product of exponentials of the same parameter $-\lambda >0$.
In the case of a perturbed model with transitions determined by \eqref{vecchiapert} we have also
\begin{equation}\label{detbal2}
\mu_N^{\lambda(\cdot)}(\xi)\Gamma_{x,y}^{\xi,\mathbb F}(j)=\mu_N^{\lambda(\cdot)}\big(\xi-j\left(\varepsilon^x-\varepsilon^y\right)\big)
\Gamma_{x,y}^{\xi-j\left(\varepsilon^x-\varepsilon^y\right),\mathbb F}(-j)
\end{equation}
provided that
\begin{equation}\label{prolampo}
\mu_N^{\lambda(\cdot)}(\xi)=\prod_z|\lambda(z)| e^{\lambda(z)\xi(z)}
\end{equation}
is the density an inhomogeneous product of exponentials and the relation
$\mathbb F(x,y)=\lambda(y)-\lambda(x)$ is satisfied.

The boundary dynamics \eqref{arpaceltica} satisfies the detailed balance condition with respect to an exponential measure with parameter $-\lambda>0$ coinciding with the one of the external source. This means that we have for a boundary site $x$
\begin{equation}\label{detbal3}
\mu_N^{\lambda(\cdot)}(\xi)\Gamma_{x}^\xi(j)=\mu_N^{\lambda(\cdot)}(\xi+j\varepsilon^x)\Gamma_{x}^{\xi+j\varepsilon^x}(-j)
\end{equation}
where $\mu_N^{\lambda(\cdot)}$ is like \eqref{prolampo}  and the value $\lambda(x)$ coincides with $\lambda$ in \eqref{arpaceltica}. In \eqref{detbal3} we called
$$
\Gamma_{x}^\xi(dj)=|\lambda| e^{\lambda(\xi(x)+j)}\chi_{[-\xi(x),+\infty)}(j)\,dj\,.
$$
By the above computations we obtain some general conditions to have a reversible KMP model. When the model is
not reversible the invariant measure is not known. According to the general result in \cite{JSTATrev} we obtain that if we fix the values of the chemical potentials of the boundary sources the model is reversible if the external field is of gradient type $\mathbb F(x,y)=\psi(y)-\psi(x)$ for a function $\psi$ such that $\psi(x)=\lambda(x)$ for $x\in \partial \Lambda_N$. A similar result can be obtained also for KMPd.

The Gaussian model \eqref{currga} satisfies the detailed balance condition with respect to a product of Gaussian distributions having the same arbitrary mean value and variance equal to $2\gamma^2$ where $\gamma^2$ is the variance of the random current across an edge.
\smallskip

When the system is in contact with sources having the same chemical potential we have that the KMP and the KMPd models are equilibrium models reversible with respect to product measures. Given a reference measure $\mu$ on $\mathbb R$, we denote by $\mu_\lambda$ the probability measure obtained inserting a chemical potential term of the form
\begin{equation}\label{nientefr}
\mu^{\lambda}(dx)=\frac{\mu(dx)e^{\lambda x}}{Z(\lambda)}\,,
\end{equation}
where $Z(\lambda)$ is the normalization constant.
The corresponding average density $\rho[\lambda]:=\int \mu_{\lambda}(dx)x=(\log Z(\lambda))'$ is increasing in $\lambda$ and we call $\lambda[\rho]$ the inverse function. In the case of the KMP model it is natural to fix $\mu(dx)=dx$ and restrict to negative values of $\lambda$. In this case $Z(\lambda)=-\lambda^{-1}=\rho[\lambda]$ and $\lambda[\rho]=-\rho^{-1}$. For the KMPd model we fix $\mu(dx)=\sum_{k=0}^{+\infty}\delta_k(dx)$ and again we restrict to negative values of $\lambda$. In this case we have $Z(\lambda)=(1-e^\lambda)^{-1}$, $\rho[\lambda]=\frac{e^\lambda}{1-e^\lambda}$ and $\lambda[\rho]=\log\frac{\rho}{1+\rho}$.

\subsection{Instantaneous current}\label{Ic}

The hydrodynamic scaling limit of diffusive systems has an explicit form  for models that are reversible and gradient. The gradient condition is written in terms of the so called instantaneous current \cite{KL99,Spohn}. The natural scaling limit for this class of processes is the diffusive one. For this reason the rates have to be multiplied by $N^2$ to get a non trivial scaling limit. Let $\xi_t$ be the random configuration of particles at time $t$ for the accelerated process. Let also $\mathcal J_t(x,y)$ be the net total amount of mass that has flown from $x$ to $y$ in the time window $[0,t]$. The instantaneous current for the bulk dynamics is defined as
\begin{equation}\label{istcc}
j_\xi(x,y):=\int \Gamma^\xi_{x,y}(dj)j\,.
\end{equation}
A simple argument \cite{Spohn} shows that
\begin{equation}\label{mart}
\mathcal J_t(x,y)-N^2\int_0^tj_{\xi_s}(x,y)\,ds\,,
\end{equation}
is a martingale. In particular \eqref{mart} has mean zero and the expected value of the current through a bond
can be obtained from the expected value of an additive functional involving the instantaneous current trough the same bond. The factor $N^2$ is due to the fact that we are speeding up the process.

For example the instantaneous current across the edge $(x,y)$ for the KMP process is given by
\begin{equation}\label{zamadel}
\int_{-\xi(y)}^{\xi(x)}\frac{jdj}{\xi(x)+\xi(y)}=\frac 12\left(\xi(x)-\xi(y)\right)\,.
\end{equation}
This computation shows that the KMP model is of gradient type. In general a model of stochastic particles on a lattice is called of gradient type \cite{KL99,Spohn} if the instantaneous current can be written as
\begin{equation}\label{gradt}
j_{\xi}(x,y)=\tau_xh(\xi)-\tau_yh(\xi)\,,
\end{equation}
where $h$ is a local function and $\tau_z$ is the shift operator by the vector $z$. Formula \eqref{zamadel} shows for example that for KMP formula \eqref{gradt} holds with $h(\xi)=\frac{\xi(0)}{2}$. Also KMPd is gradient with respect to the same function $h$.

The instantaneous current for the weakly asymmetric KMP model in the case of a constant external field $E$ in the direction from $x$ to $y$ is given by
\begin{eqnarray}\label{ordmai}
& & j_{\xi}^E(x,y)=\int_{-\xi(y)}^{\xi(x)}\Gamma^{\xi,E}_{x,y}(j)jdj\nonumber \\
& &=\frac{2}{E(\xi(x)+\xi(y))}\left[e^{\frac E2\xi(x)}\xi(x)
+e^{-\frac E2\xi(y)}\xi(y)-2\frac{e^{\frac E2\xi(x)}-e^{-\frac E2\xi(y)}}{E}\right]\nonumber  \\
& &=\frac 12\big(\xi(x)-\xi(y)\big)+\frac E6\big[\xi(x)^2+\xi(y)^2-\xi(x)\xi(y)\big]+o(E)\,.
\end{eqnarray}
The hydrodynamic behavior of the model under the action of an external field in the weakly asymmetric regime, i.e. when the external field $E$ is of order $1/N$, is determined by the first two orders in the expansion \eqref{ordmai}. In particular any perturbed KMP model having the same expansion as in \eqref{ordmai} will have the same hydrodynamic behavior of the model \eqref{vecchiapert} in the weakly asymmetric regime.

For the KMPd model with a discrete version of the above computation we get
\begin{equation}\label{ordmaid}
j_{\xi}^E(x,y)
=\frac 12\big(\xi(x)-\xi(y)\big)+\frac{E}{12}\big[2\xi(x)^2+2\xi(y)^2-2\xi(x)\xi(y)+\xi(x)+\xi(y)\big]+o(E)\,.
\end{equation}

%\begin{equation}\label{compss}
%\Gamma^E_{\xi(x),\xi(y)}(j)=\frac{1+Ej}{\xi(x)+\xi(y)}+o(E)
%\end{equation}

\section{Asymmetric models}\label{AM}

We consider now  some possible one dimensional totally asymmetric models for which the mass can move only in one preferred direction.
If on a bond $(x,y)$ the asymmetry is from $x$ to $y$ then we have that the measure $\Gamma_{x,y}^\xi$ determining the distribution
of the current has a support contained on the interval
$[0,\xi(x)]$. We consider only the KMP case and assume that the density $\Gamma_{x,y}^\xi$ depends only on $\xi(x)$ and not on $\xi(y)$.
The distribution $\Gamma_{x,y}^\xi$ of a totally asymmetric model should be obtained as a limit for large values of a constant external field of the distribution $\Gamma_{x,y}^{\xi,E}$ of a weakly asymmetric model having an instantaneous current with an expansion like \eqref{ordmai}. This is because macroscopically we will discuss the limit for large values of the field of weakly asymmetric models  having an hydrodynamic behavior deduced from the expansion \eqref{ordmai}. Since the expansion \eqref{ordmai} is for small values of the field while we discuss here microscopically the behavior for large values of the field it is reasonable to have some freedom in the determination of the limiting models. We discuss indeed two different cases. One has a product invariant measure while for the other one we discuss a duality representation of the invariant measure using a convex analytic approach.

The macroscopic domain is $\Lambda=(0,1)$ and the asymmetry is in the positive direction. For simplicity of notation we consider the models defined on the lattice $\{1,2,\dots ,N\}$ instead of $\Lambda_N$. Since the computations in this section are only microscopic the lattice size is not relevant.

\subsection{Totally asymmetric KMP model version 1}\label{tkmp1dis}

In this section we discuss a model with  distribution of the current flowing across a bond in the bulk given by
\begin{equation}\label{TAKMP1}
\Gamma_{x,x+1}^\xi= \chi_{[0,\xi(x)]}(j)\, dj\,.
\end{equation}
We fix the interaction with the boundary left source like in \cite{JSTATrev}. We imagine to have a ghost site at $0$ where an exponential random energy of parameter $-\lambda>0$ is available and this random energy available at site $0$ is transported into site
$1$ with the same mechanism \eqref{TAKMP1} of the bulk. We have then the boundary part of the dynamics at the boundary site $1$
given by
\begin{eqnarray}\label{pera}
L_{1}f(\xi) &=& \int_{0}^{+\infty}|\lambda| e^{\lambda z}\left(\int_0^{z}dj\left[f(\xi+j\varepsilon^{1})-f(\xi)\right]\right)\,dz\\
& = & \int_0^{+\infty} e^{\lambda j} \left[f(\xi+j\varepsilon^{1})-f(\xi)\right]\,dj\,.
\end{eqnarray}
At the right boundary the dynamics is like on the bulk but the mass that is moving to the right exits
from the system and disappears. More precisely at site $N$ with rate $\xi(N)$ the amount of mass present is transformed into
$\xi'(N)$ that is uniformly distributed on $[0,\xi(N)]$. We could imagine  a different mechanism that allows also a creation of mass connected with a reservoir with a given chemical potential. This mechanism however changes the distribution of the mass just at site $N$ and in particular is not  observable macroscopically.
Consider the product measure (recall $\lambda<0$)
\begin{equation}
\label{scat}
\mu_N^\lambda(d\xi)=\prod_{x\in \Lambda_N}|\lambda| e^{\lambda\xi(x)}d\xi(x)\,.
\end{equation}
With a change of variables we get
 \begin{eqnarray}\label{1}
\mathbb E_{\mu_N^\lambda}\left[L_{x,x+1}f\right] &=&\int_{ (\mathbb R^+)^N}  \mu_N^\lambda(d\xi)\int_0^{\xi(x)}dj\left[f\left(\xi-j\left(\varepsilon^x-\varepsilon^{x+1}\right)\right)-f(\xi)\right]\nonumber\\
&=&\int_{(\mathbb R^+)^N}  \mu_N^\lambda(d\xi)f(\xi)\left[\xi(x+1)-\xi(x)\right]\,.
 \end{eqnarray}
Still with a changes of variables at the boundaries we get
 \begin{eqnarray}\label{2}
\mathbb E_{\mu_N^\lambda}\left[L_{1}f\right] &=&\int_{ (\mathbb R^+)^N} \mu_N^\lambda(d\xi)\int_0^{+\infty}e^{\lambda j}\left[f(\xi+j\varepsilon^{1})-f(\xi)\right]\, dj\nonumber \\
&=&\int_{(\mathbb R^+)^N} \mu_N^\lambda(d\xi)f(\xi)\left[\xi(1)+\lambda^{-1}\right]\,,
 \end{eqnarray}
 and
 \begin{eqnarray}\label{3}
\mathbb E_{\mu_N^\lambda}\left[L_{N}f\right] &=&\int_{(\mathbb R^+)^N} \mu_N^\lambda(d\xi)\int_0^{\xi(N)}\left[f\left(\xi-j\varepsilon^{N}\right)-f(\xi)\right]\, dj\nonumber\\
&=&-\int_{(\mathbb R^+)^N} \mu_N^\lambda(d\xi)f(\xi)\left[\lambda^{-1}+\xi(N)\right]\,.
 \end{eqnarray}
Summing up \eqref{1}, \eqref{2} and \eqref{3} we obtain that \eqref{scat} is invariant for the dynamics.

\subsection{Totally asymmetric KMP version 2}

In this section we discuss a second possible totally asymmetric limit dynamics. This is the model that is obtained considering a constant external field in \eqref{vecchiapert} and taking the limit suitably normalizing the rates. The dynamics on the bulk is defined by a distribution of the current flowing across a bond given by
\begin{equation}\label{TAKMP2}
\Gamma_{x,x+1}^\xi=\delta_{\xi(x)}\,.
\end{equation}
This means that at rate one all the mass present on a site jumps to the nearest neighbor site on the right.
On the torus this dynamics is not irreducible since eventually all the mass will concentrate on a single lattice site moving randomly like an asymmetric random walk. The boundary driven case has not this problem and the dynamics is irreducible.

\smallskip

Given two probability measures $\mu$ and $\nu$ on $\mathbb R$ we define their
convolution $\mu\ast\nu=\nu\ast\mu$ as the measure on $\mathbb R$
defined by
$$
[\mu\ast\nu](A)=\int_\mathbb R \mu(A-x) d\nu(x)\,,
$$
for any measurable subset $A$. Let us also define the family of
Gamma measures $\left\{\gamma_n\right\}_{n\geq 0}$ of parameter $|\lambda|$  as follows. We
set $\gamma_0:=\delta_0$, then we define $\gamma_1$ as the absolute
continuous probability measure on $\mathbb R^+$ having density $|\lambda|
e^{\lambda x}$. Finally we define $\gamma_n:=\gamma_1^{\ast n}$
where the right hand side symbol means a n-times convolution of
$\gamma_1$. Note that $\gamma_j\ast\gamma_i=\gamma_{i+j}$.
We fix the dynamics at the boundaries like
\begin{equation}\label{casanuova}
L_{1}f(\xi)=\int_0^{+\infty}\gamma_1(j)\left[f(\xi+j\varepsilon^{1})-f(\xi)\right]\,dj
\end{equation}
and
\begin{equation}\label{casanuovaN}
L_{N}f(\xi)=\left[f\left(\xi-\xi(N)\varepsilon^{N}\right)-f(\xi)\right]\,.
\end{equation}

The invariant measure for this second version of the totally asymmetric KMP model is not of product type and it seems not to have a simple expression. We give a representation of this measure as a convex combination of products of Gamma distributions. This is done developing a kind of duality between this process and a totally asymmetric version of KMPd. It is interesting to analyze this duality within the general approach to duality in \cite{CGRS} where the case of an asymmetric KMP model is also discussed.

Consider a product measure $\nu_N$ having marginals $\nu^{(x)}$, i.e.
\begin{equation}\label{prodguess}
\nu_N(d\xi):=\prod_{x=1}^N\nu^{(x)}(d\xi(x))=:\otimes_{x=1}^N\nu^{(x)}\,.
\end{equation}
For a measure of this type we have
\begin{eqnarray}
& &\int_{\left(\mathbb R^+\right)^n}\nu_N(d\xi)\int_0^{+\infty}\gamma_k(j)f(\xi+j\varepsilon^{1})\,dj=\nonumber \\
& &\int_{\left(\mathbb R^+\right)^n}(\nu^{(1)}\ast \gamma_k)(d\xi(1))\nu^{(2)}(d\xi(2))\dots \nu^{(N)}(d\xi(N))\, f(\xi)\,.\label{una}
\end{eqnarray}
Likewise we have
\begin{eqnarray}
& &\int_{\left(\mathbb R^+\right)^n}\nu_N(d\xi)f\big(\xi +\xi(x)(\varepsilon^{x+1}-\varepsilon^x)\big)=\nonumber\\
& &\int_{\left(\mathbb R^+\right)^n}\nu^{(1)}(d\xi(1))\dots \gamma_0(d\xi(x))(\nu^{(x)}\ast\nu^{(x+1)})(d\xi(x+1))\dots \nu^{(N)}(d\xi(N))\,f(\xi)\,.\nonumber
\end{eqnarray}
The right hand side in the above formula is the expected value of the function $f$ with respect
to a product measure having $x$-marginal equal to $\gamma_0$,
$(x+1)-$marginal equal to $\nu^{(x)}\ast\nu^{(x+1)}$ and all the remaining
marginal equal to the one of $\nu_N$. Finally we have also that
\begin{eqnarray}
& &\int_{\left(\mathbb R^+\right)^n}\nu_N(d\xi)f\left(\xi-\xi(N)\varepsilon^{N}\right)=\nonumber \\
& &\int_{\left(\mathbb R^+\right)^n}\nu^{(1)}(d\xi(1))\dots
\gamma_0(d\xi(N))\,f(\xi)\,.\label{tre}
\end{eqnarray}
The above computations give that the action of the asymmetric generator $\mathcal L_{N,a}$ on a measure $\nu_N$ like in \eqref{prodguess}
is given by
\begin{eqnarray}\label{niente}
\nu_N\mathcal L_{N,a}&=& \Big\{\left[\left(\nu^{(1)}\ast \gamma_1\right)\otimes \nu^{(2)}\otimes \dots \otimes \nu^{(N)}\right]-\left[\nu^{(1)}\otimes \dots \otimes \nu^{(N)}\right]\Big\}\nonumber \\
&+& \sum_x\Big\{\left[\nu^{(1)}\otimes \dots \gamma_0\otimes \left(\nu^{(x)}\ast\nu^{(x+1)}\right)\otimes \dots \otimes \nu^{(N)}\right]-
\left[\nu^{(1)}\otimes \dots \otimes \nu^{(N)}\right]\Big\}\nonumber \\
& & +\Big\{\left[\nu^{(1)}\otimes \dots \otimes \nu^{(N-1)}\otimes \gamma_0\right] -\left[
\nu^{(1)}\otimes \dots \otimes \nu^{(N)}\right]\Big\}\,.
\end{eqnarray}
We can now show that there is a solution of the Kolmogorov equation $\partial_t\nu_N(t)=\nu_N(t)\mathcal L_{N,a}$ that can be written in the form
\begin{equation}
\nu_N(t)=\sum_{\eta\in \mathbb N^N}c_t(\eta)\gamma_{\eta(1)}\otimes\gamma_{\eta(2)}\otimes \dots
\otimes\gamma_{\eta(N)}\,.\label{guess}
\end{equation}
In the formula \eqref{guess} $\eta=(\eta(1),\dots \eta(N))\in \mathbb N^N$ can be interpreted as a configuration
of particles on the lattice and
$c_t(\eta)\geq 0$ for any fixed $t$ is a suitable probability measure on $\mathbb N^N$
to be determined. Formula \eqref{guess} says that we are searching for a
solution that can be written as a mixture of products
of Gamma measures for any time. Defining
$$
\gamma^\eta:=\gamma_{\eta(1)}\otimes\gamma_{\eta(2)}\otimes \dots
\otimes\gamma_{\eta(N)}
$$
we write compactly \eqref{guess} as
$\sum_\eta c_t(\eta)\gamma^{\eta}$. For a measure of the type \eqref{guess} we have
\begin{equation}\label{notte}
\nu_N(t)\mathcal L_{N,a}=\sum_\eta c_t(\eta)\left(\gamma^\eta\mathcal L_{N,a}\right)\,,
\end{equation}
and we can now use formula \eqref{niente}.
Reorganizing the terms, the right hand side of \eqref{notte} becomes
\begin{eqnarray}\label{giorno}
& &\sum_\eta \gamma^\eta\Big\{\left[c_t(\eta-\varepsilon^1)\chi(\eta(1)>0)-c_t(\eta)\right]\nonumber \\
& &+\sum_{x=1}^{N-1}\sum_{k=0}^{\eta(x+1)}
\left[\chi(\eta(x)=0)c_t(\eta+k(\varepsilon^{x}-\varepsilon^{x+1}))-c_t(\eta)\right]\nonumber \\
& &\left.+\left[\chi(\eta(N)=0)\sum_{k=0}^{+\infty}c_t(\eta+k\varepsilon^N)-c_t(\eta)\right]\right\}\,,
\end{eqnarray}
where $\chi$ denotes the characteristic function.
Using \eqref{giorno} we can write formula \eqref{notte} compactly as
\begin{equation}\label{stuf}
\sum_\eta\gamma^\eta\partial_t c_t(\eta)=\nu_N(t)\mathcal L_{N,a}=\sum_\eta\gamma^\eta\left(c_t(\eta)\mathcal L_{N,a}^d\right)
\end{equation}
where $\mathcal L_{N,a}^d$ is a Markov generator of a stochastic dynamics on the variables $\eta$. We interpret
\eqref{stuf} as a duality relationship between the two stochastic dynamics $\mathcal L_{N,a}$ and $\mathcal L_{N,a}^d$.
The upper index $d$ is the shorthand of \emph{dual}. The variables $\eta$ represent configurations of particles on the lattice
and $\eta(x)$ that is always an integer number is the number of particles at site $x$. By formula \eqref{giorno} the stochastic dynamics associated to $\mathcal L_{N,a}^d$ can be described as follows. In the bulk the dynamics has a distribution of current
given by $\Gamma_{x,x+1}^\eta=\delta_{\eta(x)}$. At the left boundary one particle is created with rate $1$ while all the particle at the right boundary are erased at rate $1$. This is a totally asymmetric version of the model KMPd.

We proved that the model with generator $\mathcal L_{N,a}$ starting at time zero with a distribution of the type $\sum_\eta c_0(\eta)\gamma^\eta(d\xi)$ will have a distribution of energies at time $t$ that is $\sum_\eta c_t(\eta)\gamma^\eta(d\xi)$ where $c_t(\eta)$ is the distribution of particles at time $t$ for the model with generator $\mathcal L_{N,a}^d$ starting at time $0$ with the distribution of particles given by $c_0(\eta)$. In particular, considering the limit for $t\to +\infty$, this relationship between the two processes will hold also for the corresponding invariant measures for which we get
\begin{equation}\label{dual-inv}
\mu_N(d\xi)=\sum_\eta \mu_{N,d}(\eta)\gamma^\eta(d\xi)\,.
\end{equation}
In \eqref{dual-inv} $\mu_N$ is the invariant measure for the process $\mathcal L_{N,a}$ while $\mu_{N,d}$ is the invariant measure for the process $\mathcal L^d_{N,a}$.

\section{Scaling limits}

\subsection{Scaling limit}
The KMP model is gradient and the hydrodynamic behavior is relatively well understood \cite{BGL,KL99,Spohn}. We consider the one dimensional case with $\Lambda=(0,1)$. In the symmetric case the model is diffusive and the natural scaling of the system is obtained considering a lattice of mesh $\frac 1N$ and rescaling time by a factor $N^2$. This is done simply multiplying by $N^2$ the rates of jump (for notational convenience we will multiply by a factor $2N^2$). The observable that describe macroscopically the evolution of the mass of the system is the so called empirical measure. This is a positive measure on $\Lambda$ associated to any fixed microscopic configuration $\xi$. It is defined as a convex combination of delta measures as
\begin{equation}\label{empmisxi}
\pi_N(\xi):=\frac 1N\sum_{x\in \Lambda_N}\xi(x)\delta_x\,.
\end{equation}
Integrating a continuous function $f:\Lambda\to \mathbb R$ with respect to $\pi_N(\xi)$ we get
\begin{equation}\label{defpif}
\int_\Lambda f\,d\pi_N(\xi)=\frac 1N\sum_{x\in \Lambda_N}f(x)\xi(x)\,.
\end{equation}
In the hydrodynamic scaling limit the empirical measure, that for any finite $N$ is atomic and random, becomes deterministic and absolutely continuous. For suitable initial conditions $\xi_0$ that are associated to a given density profile $\gamma(x)dx$ in the sense that
\begin{equation}\label{asc}
\lim_{N\to +\infty}\int_\Lambda f\, d\pi_N(\xi_0)=\int_\Lambda f(x)\gamma(x)dx
\end{equation}
we have that $\pi_N(\xi_t)$ is associated to the density profile $\rho(x,t)dx$ where $\rho$ is the solution to the heat equation
with initial condition $\gamma$. The boundary conditions are fixed by the interactions with the external sources. We consider the case when the boundary dynamics is also accelerated by $N^2$ and the final effect of this fast interaction is that the values of the densities at the boundaries are fixed by the external sources \cite{ELS}. We have then that $\rho$ is the solution of a Cauchy problem with Dirichelet boundary condition like
\begin{equation}\label{drc}
\left\{
\begin{array}{l}
\partial_t\rho=\Delta \rho\\
\rho(x,0)=\gamma(x)\\
\rho(0,t)=\rho_-\\
\rho(1,t)=\rho_+\,.
\end{array}
\right.
\end{equation}
Without loss of generality we will always consider the case $\rho_-\leq \rho_+$.
This is a space time law of large numbers and the corresponding fluctuations can be described by a large deviations principle \cite{BGL}. Given a space and time dependent density profile $\rho(x,t)dx$ the probability that the empirical measure will be in a suitable neighborhood of it is exponential unlikely with a corresponding rate functional that we call
dynamic large deviations rate functional. This is the main ingredient for the dynamic variational study of stationary non equilibrium states of the macroscopic fluctuation theory \cite{MFT}.

To show the dynamic large deviations rate functional we have before to discuss the hydrodynamic behavior of a weakly asymmetric version of the model. This has the bulk part of the generator obtained as a sum of possibly time dependent contributions like \eqref{bulkxyE} multiplied by $N^2$. In particular we consider a model with rates determined by \eqref{vecchiapert} or more generally
having an expansion like \eqref{ordmai} where the external field $E$ is however substituted by a space and time dependent vector field $\mathbb F$  obtained by a discretization of a smooth vector field on $\Lambda$ like \eqref{disc-vec}.
The hydrodynamic behavior of this model is similar to the symmetric one and the external field appears macroscopically with a new term. The hydrodynamic equation is
\begin{equation}\label{drcE}
\left\{
\begin{array}{l}
\partial_t\rho(x,t)=\Delta \rho(x,t)-\nabla \cdot \left(\rho^2(x,t)F(x,t)\right)\,,\\
\rho(x,0)=\gamma(x)\,,\\
\rho(0,t)=\rho_-\,,\\
\rho(1,t)=\rho_+\,.
\end{array}
\right.
\end{equation}

\subsection{Transport coefficients}
The general form of the hydrodynamic equation associated to weakly asymmetric diffusive stochastic particle systems is
\begin{equation}\label{hydrog}
\partial_t\rho=\nabla \cdot\left(D(\rho)\nabla \rho-\sigma(\rho)F\right)\,.
\end{equation}
The symmetric and positive definite matrix $D$ is called the diffusion matrix while the symmetric and positive definite matrix $\sigma$ is called the mobility.
For all the models that we are discussing the diffusion matrix coincides with the identity matrix while the mobility
is a multiple of the identity matrix $\sigma(\rho)\mathbb I$ (we are calling $\sigma$ both the matrix and the scalar value on the diagonal). The transport coefficients can be explicitly computed for gradient models for which at  equilibrium the invariant measure is known.

It is convenient to write the hydrodynamic equation \eqref{hydrog} as a conservation law $\partial_t\rho+\nabla\cdot J_F(\rho)=0$ where
\begin{equation}\label{jeff}
J_F(\rho):=-\nabla \rho+\sigma(\rho)F
\end{equation}
is the typical current observed in correspondence to the density profile $\rho$.

We briefly illustrate the general structure of the proof of the hydrodynamic limit for gradient reversible models that
allows to identify and compute exactly the transport coefficients \cite{KL99,S}. This argument is the bridge between the microscopic and the macroscopic description of a system and allows to identify the hydrodynamic equations associated to the microscopic models.

The starting point for the hydrodynamic description of the system is the discrete continuity equation that is
\begin{equation}\label{disc-cont}
\xi_t(x)-\xi_0(x)=-\nabla\cdot \mathcal J_t(x)\,,
\end{equation}
where $\mathcal J_t$ has been defined in section \ref{Ic} and $\nabla\cdot$ denotes the discrete divergence. This is defined for a discrete vector field $\phi$ as $\nabla\cdot\phi(x):=\sum_{y\sim x}\phi(x,y)$.
Using \eqref{mart} we can rewrite \eqref{disc-cont} as
\begin{equation}\label{rev}
\xi_t(x)-\xi_0(x)= -N^2\int_0^t\nabla\cdot j_{\xi_s}(x)\,ds+ M_t(x)\,,
\end{equation}
where $M_t(i)$ is a martingale term obtained by summing some martingales of the type \eqref{mart}.

Since we are interested only on the transport coefficients we consider the model defined on the 1-dimensional continuous torus $\Lambda=[0,1]$ with periodic boundary conditions.
Multiplying equalities \eqref{rev} by a test function $\psi$, dividing by $N$ and summing over $x$  we obtain
\begin{equation}\label{gate}
\int_{[0,1]} \psi \,d\pi_N(\xi_t)-\int_{[0,1]} \psi\, d\pi_N(\xi_0)=-N\int_0^t\sum_x\nabla\cdot j_{\xi_s}(x)\,\psi\left(x\right)\,ds+o(1)\,.
\end{equation}
The infinitesimal term comes from the martingale terms and can be shown
to be negligible in the limit of large $N$ \cite{KL99, Spohn}. Using the gradient condition \eqref{gradt} and  performing a double discrete integration by part, up to the infinitesimal term one has that the right hand side of \eqref{gate} is
\begin{equation}\label{aibp}
\frac{1}{N}\sum_x\int_0^t \tau_xh(\xi)\left[N^2\left(\psi\left(x+\frac{1}{N}\right)+\psi\left(x-\frac{1}{N}\right)-
2\psi\Big(x\Big)\right)\right]\,ds\,.\nonumber \\
\end{equation}
Considering a $C^2$ test function $\psi$ the term inside squared parenthesis in the last term of \eqref{aibp}
coincides with $\Delta\psi\left(x\right)$ up to an uniformly infinitesimal term.

At this point the main issue in proving hydrodynamic behavior is the prove the validity of a local equilibrium property. Let us define
\begin{equation}\label{local}
A(\rho)=\mathbb E_{\mu_{N}^{\lambda[\rho]}}\left(h (\xi)\right)\,,
\end{equation}
where  $\mu_N^\lambda$ is the product of measures \eqref{nientefr} associated to the chemical potential $\lambda$ and we recall that
$\lambda[\rho]$ is the chemical potential associated to the density $\rho$ (see the discussion after \eqref{nientefr}).
The local equilibrium property is explicitly stated through a replacement lemma that states that
\begin{equation}\label{repla}
\frac{1}{N}\sum_x\int_0^t \tau_xh(\xi)\Delta\psi\left(x\right)\,ds\simeq
\frac{1}{N}\sum_x\int_0^t A\left(\frac{\int_{B_{x}}d\pi_N(\xi_s)}{|B_{x}|}\right)\Delta\psi\left(x\right)\,ds
\end{equation}
where $B_{x}$ is a microscopically large but macroscopically small volume around
the point $x\in \Lambda_N$.
This allows to write, up to infinitesimal corrections, equation \eqref{gate} in terms only of the empirical measure.
Substituting the r.h.s. of \eqref{repla} in the place of the r.h.s. of \eqref{gate}, we obtain that in the limit of large $N$ the empirical measure $\pi_N(\eta_t)$ converges to $\rho(x,t)dx$ satisfying for any smooth test function $\psi$
\begin{equation}\label{widro}
\int_{0}^1 \psi(x)\rho(x,t)\,dx-\int_{0}^1\psi(x)\rho(x,0)\,dx=\int_0^t ds\int_{0}^1 A(\rho(x,s))\Delta\psi(x)\,dx\,.
\end{equation}
equation \eqref{widro} is a weak form of the hydrodynamic equation \eqref{hydrog} with $F=0$
and having a diagonal diffusion matrix having each term in the diagonal equal to
\begin{equation}\label{diff}
D(\rho)= \frac{ d A(\rho)}{d \rho}\,.
\end{equation}
 For a mathematical discussion of this issue see \cite{KL99} Chapter 5. For all the models that we are discussing we have that $h(\xi)=\frac{\xi(0)}{2}$ so that $A(\rho)=\frac{\rho}{2}$. For notational convenience in order to have an unitary diffusion matrix we multiply all the rate of transition by a factor of 2 and correspondingly the diffusion matrix is the identity matrix.

\smallskip

We show now a computation that allows to determine the mobility of the models from the microscopic dynamics.  Let us consider weakly asymmetric models subject to an external field obtained by the discretization \eqref{disc-vec} of a smooth vector field $F$.
Consider the time window $[0,t]$ and we still speed up the process by a $N^2$ factor. The scalar product of the flow of mass in this time window with a vector field $H$ is given by
\begin{equation}\label{total-work}
\frac {1}{N^d}\sum_{x\sim y}\mathcal J_t(x,y)\mathbb H (x,y)\,.
\end{equation}
In formula \eqref{total-work} the sum is over unordered nearest neighbor sites. By the antisymmetry of the two vector fields there is no ambiguity in this definition. The factor $N^{-d}$ is due to the fact that the scaling limit normalizes the mass by this factor.
Using \eqref{mart} and \eqref{ordmai} we can write \eqref{total-work} up to a neglecting martingale term as
\begin{equation}\label{work-up}
N^{2-d}\int_0^t\sum_{x\sim y}j^{\mathbb F}_{\xi_s}(x,y)\mathbb H(x,y)\,ds\,.
\end{equation}
For simplicity we consider the one dimensional KMP model with a macroscopic constant external field $F$. Microscopically this corresponds to consider $E=\frac{F}{N}$ in \eqref{ordmai}. The value $\frac FN$ is indeed the discretized value for a lattice of size $1/N$ corresponding to a constant macroscopic external field $F$.
We introduce the function
\begin{equation}\label{func-g}
g(\eta)=\frac 16\big(\xi(0)^2+\xi(1/N)^2-\xi(0)\xi(1/N)\big)
\end{equation}
that, suitably shifted, multiplies $E$ in the right hand side of \eqref{ordmai}. In the case of KMPd we have instead to use formula \eqref{ordmaid}.
We can write \eqref{work-up} in $d=1$ as
\begin{equation}
N\int_0^t\sum_{x}j_{\xi_s}\left(x,x+\frac 1N\right)\mathbb H\left(x,x+\frac 1N\right)ds+ F\int_0^t\sum_{x}\tau_xg(\xi_s)\mathbb H\left(x,x+\frac 1N\right)ds\,.
\label{primo-passo}
\end{equation}
Since the current $j$ is gradient, recalling \eqref{gradt}, formula \eqref{primo-passo} becomes after a discrete integration by parts
\begin{equation}
N\int_0^t\sum_{x}\tau_x h(\xi_s)\nabla\cdot \mathbb H(x)\,ds+ F \int_0^t\sum_{x}\tau_xg(\xi_s)\mathbb H\left(x,x+\frac 1N \right)\,ds\,.
\label{secondo-passo}
\end{equation}
Since the vector field $H$ is smooth, recalling its definition \eqref{disc-vec},
we have that $N^2\nabla\cdot \mathbb H(x)=\nabla\cdot H(x)$ up to uniform  infinitesimal terms. Applying also in this case
the replacement Lemma we have that with high probability when $N$ is diverging \eqref{secondo-passo} converges to
\begin{equation}\label{cenone}
\int_0^t ds\int_{0}^1dx\,\left[A(\rho(x,s))\nabla\cdot H(x)+F\sigma(\rho(x,s))H(x)\right]\,,
\end{equation}
where
\begin{equation}\label{finalmente}
\sigma(\rho)=\mathbb E_{\mu^{\lambda[\rho]}_N}\left[g(\eta)\right]\,.
\end{equation}
Formula \eqref{cenone} is a weak form of $\int_0^tds\int_0^1J_F(\rho)\cdot H\,dx$ with
\begin{equation}\label{tipF}
J_F(\rho)=-D(\rho)\nabla\rho +\sigma(\rho)F\,.
\end{equation}
This is the typical current associated to a density profile $\rho$ in presence of an external field $F$.
Recall that for notational convenience we are multiplying the rates by a factor of 2 so that formula \eqref{finalmente}
for ours prototype models gives
\begin{equation}\label{lesigma}
\sigma(\rho)=\left\{
\begin{array}{ll}
\rho^2 & \textrm{KMP}\\
\rho(\rho+1) & \textrm{KMPd}\\
\rho^2+1 & \textrm{KMPx}\,.
\end{array}
\right.
\end{equation}

\smallskip

A general identity holding for diffusive systems is the Einstein relation between the transport coefficients and the density of free energy \cite{MFT,Spohn}
\begin{equation}\label{Einzwei}
D(\rho)=\sigma(\rho)f''(\rho)\,.
\end{equation}
In \eqref{Einzwei} $f$ is the density of free energy that will be introduced and discussed in Section \ref{trc}. We have that $f'(\rho)=\lambda[\rho]$ where we recall $\lambda[\cdot]$ is the chemical potential as a function of the density introduced in Section \ref{unno}. The Einstein relation can be then written equivalently as
\begin{equation}\label{llaa}
D(\rho)=\sigma(\rho)\lambda'[\rho]\,.
\end{equation}
\section{Stationary solutions and currents}

The stationary solution $\bar\rho_E$ of the hydrodynamic equation \eqref{hydrog} with a constant external field $E$, in one dimension with boundary
conditions $\rho_\pm$ is obtained as the solution of
\begin{equation}\label{steq}
\left\{
\begin{array}{l}
\Delta\rho-E\nabla\sigma\left(\rho\right)=0\,, \\
\rho(0)=\rho_-\ , \rho(1)=\rho_+\,.
\end{array}
\right.
\end{equation}
Recalling the typical current \eqref{jeff}, equation \eqref{steq} can be written as $\nabla\cdot J_E(\rho)=0$. In one dimension this implies that the typical current in the stationary state is spatially constant.
We are interested in the asymptotic behavior of the stationary solution of the hydrodynamic equation in the limit of a large external field. The asymptotic behavior of the solution can be obtained either by a direct computation or using the general theory (\cite{S} chapter 15). According to this the limiting value is the stationary solution of the conservation law obtained removing the second order derivative term and with Bardos Leroux N\'{e}d\'{e}lec boundary conditions. This should be also the stationary solution of the hydrodynamic equation for the asymmetric models discussed in Section \ref{AM} \cite{Bah1,PS}. For our aims it is enough a weaker result. We use the fact that the unique solution of \eqref{steq} is monotone and the asymptotic behavior of the current for large fields.

Equation \eqref{steq} can be integrated obtaining
\begin{equation}\label{ei}
\nabla\rho-E\sigma(\rho)=-J_E
\end{equation}
where $J_E$ is the integration constant that coincides with $J_E(\bar\rho_E)$ the typical current in the stationary state.

The monotonicity of $\bar\rho_E$ follows by the fact that if there is a non constant solution $\tilde\rho$ of the equation in \eqref{steq} such that $\nabla\tilde\rho(y)=0$ for some $y\in[0,1]$ then we have two different solutions to the Cauchy problem determined by the conditions $\nabla\rho(y)=0$ and $\rho(y)=\tilde\rho(y)$. One is $\tilde\rho$ itself and the other one is the constant one.

Since the solution $\rho$ in \eqref{ei} is monotone the integration constant $J_E $ is determined imposing the validity of the boundary conditions by
\begin{equation}\label{condju}
\int_{\rho_-}^{\rho_+}\frac{d\rho}{E\sigma(\rho)-J_E}=1\,.
\end{equation}
The left hand side of \eqref{condju} is monotone on $J_E$ that can be uniquely fixed for any choice of $\rho_\pm$ and $E$. Once $J_E$ has been fixed $\bar\rho_E$ is uniquely obtained by a direct integration of \eqref{ei}.
We distinguish the stationary states according to the sign of the stationary current. For any choice of $\rho_\pm$ there exists an external field $E^*$ for which the typical value of the current in the stationary state vanishes. This field is obtained selecting $J_{E^*}=J_{E^*}(\bar\rho_{E^*})=0$ in \eqref{condju} and using \eqref{llaa}
\begin{equation}\label{estar}
E^*=\lambda[\rho_+]-\lambda[\rho_-]\,.
\end{equation}
As the intuition suggests if we have a field $E>E^*$ then $J_E(\bar\rho_E)>0$ while $J_E(\bar\rho_E)<0$ for a field $E<E^*$.

\smallskip

To study the asymptotic behavior of the current for large fields it is convenient to introduce the variable $\alpha =\frac 1E$ and the function $\mathcal E(\alpha):=\alpha J_{\frac 1\alpha}$. Condition \eqref{condju} becomes
\begin{equation}\label{condjue}
\alpha\int_{\rho_-}^{\rho_+}\frac{d\rho}{\sigma(\rho)-\mathcal E(\alpha)}=1\,.
\end{equation}
For any  $\alpha\neq 0$ the value $\mathcal E(\alpha)$ cannot belong to the interval $\{\sigma(\rho)\,, \rho\in[\rho_-,\rho_+]\}$ because otherwise the integral on the left hand side of \eqref{condjue} is divergent. Moreover when $\alpha <0$
then we need to have $\mathcal E(\alpha)\geq \sigma(\rho)$ for any $\rho$ while if $\alpha>0$ we get $\mathcal E(\alpha)\leq \sigma(\rho)$ for any $\rho$. This follows by the fact that otherwise the sign of the integral in \eqref{condjue} is not positive. When $|\alpha|\to 0$ the value of $\mathcal E(\alpha)$ cannot stay far from the interval $\{\sigma(\rho)\,, \rho\in[\rho_-,\rho_+]\}$ since otherwise the equality \eqref{condjue} cannot be satisfied. Since depending on the sign of $\alpha$ we have that $\mathcal E(\alpha)$ is always above or below the interval we deduce that
\begin{equation}\label{bgn}
\left\{
\begin{array}{l}
\lim_{\alpha\uparrow 0} \mathcal E(\alpha)=\lim_{E\to -\infty}J_E/E=\max_{\rho\in[\rho_-,\rho_+]}\sigma(\rho)\,, \\
\lim_{\alpha\downarrow 0} \mathcal E(\alpha)=\lim_{E\to +\infty}J_E/E=\min_{\rho\in[\rho_-,\rho_+]}\sigma(\rho)\,.
\end{array}
\right.
\end{equation}

\section{Dynamic large deviations and quasi-potential}
The dynamic rate functional can be described as follows. Consider the class of perturbations obtained adding an external field that is the gradient of a potential assuming the value zero at the boundaries. This means $F(x,t)=\nabla H(x,t)$ with $H(0,t)=H(1,t)=0$. Given a space time dependent density profile $\rho(x,t)dx$ we compute the potential $H$ solving the equation
\begin{equation}\label{equaH}
\left\{
\begin{array}{l}
\partial_t\rho(x,t)=\Delta \rho(x,t)-\nabla \cdot \left(\sigma(\rho(x,t))\nabla H(x,t)\right)\\
H(0,t)=H(1,t)=0\,.
\end{array}
\right.
\end{equation}
The dynamic rate function for a symmetric model in the time window $[0,T]$ is then obtained by \cite{BFG,KL99,MFT}
\begin{equation}\label{drf}
I_{[0,T]}(\rho)=\frac 14\int_0^Tdt\int_\Lambda dx\, \sigma (\rho)\left(\nabla H\right)^2\,,
\end{equation}
if the density profile satisfies the boundary conditions $\rho(x,t)=\rho_-$ and $\rho(x,t)=\rho_+$, while instead is identically equal to $+\infty$ if the boundary conditions are violated.

If the original process for which we want to compute large deviations is not the symmetric one but is already a weakly asymmetric one with for example a constant external field then the dynamic rate functional has still the form \eqref{drf} but the potential $H$ has to be computed using the equation
\begin{equation}\label{equaEH}
\left\{
\begin{array}{l}
\partial_t\rho(x,t)=\Delta \rho(x,t)-\nabla \cdot \left(\sigma(\rho(x,t))\left(E+\nabla H(x,t)\right)\right)\\
H(0,t)=H(1,t)=0\,.
\end{array}
\right.
\end{equation}
The quasi-potential $W_E$ \cite{MFT,FW} associated to a dynamic rate functional like \eqref{drf}, for a model having a constant external field $E$, is defined through the following variational problem
\begin{equation}\label{qp}
W_E(\rho):=\inf_{T>0}\inf_{\hat\rho\in\mathcal A_{\rho,T}} I_{[-T,0]}\left(\hat\rho\right)\,.
\end{equation}
In the above equation $\rho=\rho(x)dx$ is a space dependent density, the lower index $E$ denotes the external field and $\hat\rho$ is a space and time dependent density profile belonging to the set of space time profiles
\begin{equation}\label{aaa}
\mathcal A_{\rho,T}:=\left\{\hat\rho\,:\, \hat\rho(x,-T)=\bar\rho_E\,,\, \hat\rho(x,0)=\rho(x)\right\}\,.
\end{equation}

\smallskip

The problem we are interested in is the computation of the large deviation rate functional for the empirical measure when the mass is distributed according to the invariant measure $\mu_{N,E}$. In general the computation of the invariant measure in the non reversible case is a difficult problem. We give a description of the invariant measure at a large deviations scale. This asymptotic is described by the associate rate functional by
\begin{equation}\label{ldpinvm}
\mathbb P_{\mu_{N,E}}(\pi_N(\eta)\sim \rho(x)dx)\simeq e^{-NV_E(\rho)}\,.
\end{equation}
Again we denote by a lower index the dependence on the external vector field.

Under general conditions \cite{BG,B,FW} the large deviations rate functional for the invariant measure $V_E$ and the quasi-potential $W_E$ coincide $V_E(\rho)=W_E(\rho)$. We can then compute the large deviations asymptotic of the invariant measure solving the dynamic variational problem \eqref{qp} without entering into the details of the invariant measure \cite{MFT}.

\section{Quasi-potential for weakly asymmetric models}

In \cite{BGL} it was shown that for all the boundary driven one dimensional
symmetric models having constant diffusion and quadratic mobility it is possible to compute the corresponding non local quasi-potential. We show that this is possible for the same class of models also in the case of a weak constant asymmetry. The corresponding quasi-potential is still non local and has a structure similar to the one of weakly asymmetric exclusion \cite{CPAM,BGLa,ED}.

\subsection{The reversible case}\label{trc}
In the case of reversible models (see section \ref{unno}) the computation of the quasi-potential is direct and we do
not need to solve the variational problem \eqref{qp}. This is due to the fact that the invariant measure is product and the corresponding large deviations rate functional can be computed directly as the Legendre transform of a scaled cumulant generating function \cite{SH,T}. Let us show this computation in a more general framework. Consider a family of probability measures $\mu^\lambda$
on $\mathbb R$ depending on the real parameter $\lambda$ of the form \eqref{nientefr}. We call
\begin{equation}\label{prr}
P_\lambda(\phi)=\log \int_{\mathbb R}\mu^\lambda(d x)e^{\phi x}
\end{equation}
its cumulant  generating function. Using the expression \eqref{nientefr} we have that
\begin{equation}\label{ca}
P_\lambda(\phi)=P(\phi+\lambda)-P(\lambda)
\end{equation}
where $P(\cdot)=\log Z(\cdot)$. We call
\begin{equation}\label{llp}
f_\lambda(\alpha):=\sup_{\phi}\left\{\alpha\phi-P_\lambda(\phi)\right\}
\end{equation}
the Legendre transform of $P_\lambda$. Using \eqref{ca} we obtain
\begin{equation}\label{serio}
f_\lambda(\rho)=f(\rho)+P(\lambda)-\lambda\rho\,,
\end{equation}
where $f(\cdot)$ is the Legendre transform of $P$ and, in the case of models with equilibrium product measures, it is called the density of free energy. Recall that $\rho[\lambda]$ and $\lambda[\rho]$ are the monotone functions determining respectively the density as a function of the chemical potential and the chemical potential as a function of the density. We have that $\lambda[\rho]=f'(\rho)$, $P'(\lambda)=\rho[\lambda]$. By the Legendre duality
we obtain
\begin{equation}\label{freng}
f_\lambda(\rho)=f(\rho)-f\big(\rho[\lambda]\big)-f'\big(\rho[\lambda]\big)\big(\rho-\rho[\lambda]\big)\,.
\end{equation}
The density of free energy $f$ satisfies the Einstein relation \eqref{Einzwei} and we obtain
\begin{equation}\label{lef}
f(\rho)=\left\{
\begin{array}{ll}
-\log\rho & \textrm{KMP}\,,\\
\rho\log\rho-(1+\rho)\log(1+\rho) & \textrm{KMPd}\,,\\
\rho\arctan\rho-\frac 12\log(1+\rho^2) & \textrm{KMPx}\,.
\end{array}
\right.
\end{equation}
Consider a slowly varying product measure
$\mu_N^{\lambda(\cdot)}=\prod_{x\in\Lambda_N}\mu^{\lambda(x)}(d\xi(x))$ where $\lambda(\cdot)$ is a continuous function on $\Lambda$. Let $f:\Lambda\to \mathbb R$
be a continuous test function. We can compute
\begin{eqnarray}\label{ep}
&& \mathcal P(f)=\lim_{N\to +\infty}\frac{1}{N^d}\log \int_{\mathbb R^{\Lambda_N}}\prod_{x\in\Lambda_N}\mu^{\lambda(x)}(d\xi(x))
e^{N^d\int_{\Lambda} f d\pi_N(\xi)}\nonumber \\
& &=\lim_{N\to +\infty}\frac{1}{N^d}\sum_{x\in \Lambda_N}P_{\lambda(x)}(f(x))=\int_\Lambda P_{\lambda(x)}(f(x))dx\,.
\end{eqnarray}
A general theorem \cite{SH,T} implies that the large deviations rate functional $\mathcal V_{\lambda}$ for $\pi_N(\xi)$ when the configuration is distributed according to the slowly varying product measure with marginals $\mu^{\lambda(x)}$ is
obtained as the Legendre transform of \eqref{ep}
\begin{eqnarray}\label{compiti}
& &\mathcal V_{\lambda}(\rho)=\sup_{f}\left\{\int_\Lambda \rho(x)f(x)dx-\mathcal P(f)\right\}=\int_\Lambda f_{\lambda(x)}(\rho(x))dx\nonumber\\
& &=\int_\Lambda \Big[f(\rho(x))-f\big(\rho[\lambda(x)]\big)-f'\big(\rho[\lambda(x)]\big)\big(\rho(x)-\rho[\lambda(x)]\big)\Big]\,dx\,.
\end{eqnarray}
Consider a model having rates of transition \eqref{gammae} where $\mathbb F$ is the discretization of $F=\nabla \psi$ with $\psi$ is a smooth function on the domain $\Lambda$ such that $\psi|_{\partial \Lambda_N}=\lambda$.
By the general result on non homogeneous reversible models in Section \ref{unno} we have that the invariant measure is of product type slowly varying and indeed coincides with $\mu_N^{\psi(\cdot)}$. The quasi-potential can be computed using the approach above described and we have $V_{\nabla \psi}(\rho)=\mathcal V_\psi(\rho)$.

\subsection{An Hamilton-Jacobi equation}

From the general theory \cite{MFT} associated to the variational problem \eqref{qp} there is an infinite dimensional Hamilton-Jacobi equation that in this specific case is written as
\begin{equation}\label{HJ}
\int_{\Lambda}\left[\nabla \frac{\delta V(\rho)}{\delta \rho}\cdot \sigma(\rho)\nabla\frac{\delta V(\rho)}{\delta \rho}-
\frac{\delta V(\rho)}{\delta \rho}\nabla\cdot\Big(\nabla \rho-E \sigma(\rho)\Big)\right]dx=0\,.
\end{equation}
In the above formula $\frac{\delta V(\rho)}{\delta \rho}$ denotes the functional derivative.
The rate functional $V_E$ that coincides with the quasi-potential $W_E$ is a solution to \eqref{HJ}.
We show that it is possible to find the relevant solutions of this equation for all the weakly asymmetric one dimensional models discussed here. The cases with zero external field unitary diffusion matrix and quadratic mobility were discussed in \cite{BGL}. The symmetric and weakly asymmetric cases with unitary diffusion matrix and quadratic concave mobility correspond essentially to exclusion models and have been already discussed in \cite{CPAM, BGLa, ED}. Here we complete the class of solvable models discussing the cases of unitary diffusion, quadratic and convex mobilities and in presence of a constant external field $E$. A similar computation for a model of oscillators having the same dynamic rate functional as the KMP model has been done in \cite{BP}. In \cite{bpc} it is considered the KMP case and its totally asymmetric limit.

We show in this section how to find solutions of the Hamilton-Jacobi equation \eqref{HJ}. Later on we discuss
more precisely the relevant variational problems corresponding to the different values of the external field. This second step is not discussed in full detail since a complete analysis  requires a long discussion. The variational problems are however very similar to the corresponding ones for the exclusion process and we refer to \cite{Lag,CPAM,BGLa} for the details.
Following the approach of \cite{BDGJL1,BDGJL2,MFT, BGL} we search for a solution of \eqref{HJ} of the form
\begin{equation}\label{theform}
\frac{\delta V(\rho(x))}{\delta \rho}=f'(\rho(x))-f'(\phi(x))\,,
\end{equation}
where $\phi$ has to be determined by the equation \eqref{HJ}. By the general theory \cite{MFT} $\phi$ has to satisfies the boundary conditions $\phi(0)=\rho_-$, $\phi(1)=\rho_+$.
We write the generic quadratic mobility as $\sigma(\rho)=c_2\rho^2+c_1\rho+c_0$
for suitable constants $c_i$. We consider only the cases with $c_2\neq 0$. The cases $c_2=0$ correspond to special models (zero range, Ginzburg-Landau) that have a local quasi-potential and can be studied directly.
We Insert \eqref{theform} into \eqref{HJ} and use the quadratic expression of the mobility with  some manipulations like in \cite{BDGJL1,BDGJL2,MFT,BGL}. After one integration by parts whose boundary terms disappear since $\rho$ and $\phi$ satisfy the same boundary conditions we obtain
\begin{equation}\label{primoc}
\int_\Lambda\left[\nabla\big(f'(\phi)-f'(\rho)\big)\sigma(\rho)\nabla f'(\phi)\right]\,dx +E\int_\Lambda\left(f'(\phi)-f'(\rho)\right)\nabla \sigma(\rho)\,dx=0\,.
\end{equation}
The first term in \eqref{primoc} can be developed as follows.
First we compute the derivatives and add and subtract suitable terms getting
\begin{equation}\label{secondoc}
\int_\Lambda\left[\nabla\left(\phi-\rho\right)\frac{\nabla\phi}{\sigma(\phi)}+
\left(\sigma(\rho)-\sigma(\phi)\right)\left(\frac{\nabla\phi}{\sigma(\phi)}\right)^2\right]\,dx\,.
\end{equation}
Then we integrate by parts the first term in \eqref{secondoc} and use the identity
\begin{equation}\label{formula}
\sigma(\rho)-\sigma(\phi)=(\rho-\phi)(c_2(\rho+\phi)+c_1)
\end{equation}
obtaining
\begin{equation}\label{terzoc}
\int_\Lambda\left[\frac{\rho-\phi}{\sigma(\phi)}\Delta\phi+c_2\left(\frac{\rho-\phi}{\sigma(\phi)}\right)^2\left(\nabla \phi\right)^2\right]\, dx\,.
\end{equation}
For the second term in \eqref{primoc} we integrate by parts and use again \eqref{formula} obtaining
\begin{equation}\label{quartoc}
E\int_\Lambda\left[\frac{(\phi-\rho)\nabla\phi}{\sigma(\phi)}\big(c_2(\rho+\phi)+c_1\big)\right]\,dx\,.
\end{equation}
Putting together \eqref{terzoc} and \eqref{quartoc} we obtain that the Hamilton Jacobi equation can be written as
\begin{equation}\label{onestep}
\int_{\Lambda}\frac{(\rho-\phi)}{\sigma^2(\phi)}\Big[\Delta\phi\sigma(\phi)+c_2(\nabla\phi)^2(\rho-\phi)-
E\sigma(\phi)\nabla\phi\left(c_2(\rho+\phi)+c_1\right)\Big]dx=0\,.
\end{equation}
A possible way of solving the above equation is to impose that the term inside squared parenthesis is zero. Let us introduce the following functional on  $\rho,\phi$
\begin{equation}\label{effeciu}
\mathcal G_E(\rho,\phi):=\int_\Lambda\Big[f(\rho)-f(\phi)- f'(\phi)(\rho-\phi)\Big]dx + \mathcal R(\phi)\,,
\end{equation}
where
\begin{equation}\label{arrr}
\mathcal R(\phi)=\int_\Lambda \frac{1}{c_2E\sigma(\phi)}\Big[(\nabla\phi-E\sigma(\phi))\log|\nabla\phi-E\sigma(\phi)|-\nabla\phi\log|\nabla\phi|\Big]\,.
\end{equation}
The case $E=0$ in \cite{BGL} can be recovered as a limit.
Observe that the first term in \eqref{effeciu} corresponds to an equilibrium rate functional with typical density profile $\phi$ (see \eqref{compiti}) and we have to add the new term \eqref{arrr}.
We have for the functional \eqref{effeciu} that $\frac{\delta \mathcal G_E}{\delta \rho}=f'(\rho)-f'(\phi)$ while instead with a long but straightforward computation we have
\begin{equation}\label{perfi}
\frac{\delta \mathcal G_E}{\delta \phi}=\frac{\phi-\rho}{\sigma(\phi)}-\frac{\Delta\phi}{c_2\nabla\phi(\nabla\phi-E\sigma(\phi))}
+\frac{E(c_2\phi+c_1)}{c_2(\nabla\phi-E\sigma(\phi))}\,.
\end{equation}
If the right hand side of \eqref{perfi} is zero then also the term inside the squared parenthesis in
\eqref{onestep} is zero. Let us call  $\phi^\rho$  a critical point satisfying $\frac{\delta \mathcal G_E(\rho,\phi^\rho)}{\delta \phi}=0$. We obtain that the functional of $\rho$ defined by $\mathcal G_E(\rho, \phi^\rho)$ solves the Hamilton Jacobi equation \eqref{HJ}. Since in general the critical points are not unique we discuss more in detail the specific identification of the relevant $\phi^\rho$ that gives the quasi-potential for our models. We distinguish the 3 cases $E=E^*$, $E<E^*$ and $E>E^*$.

\subsection{The case $E=E^*$}
This is a special case that corresponds to a model that has no current in the stationary state $J_{E^*}(\bar\rho_{E^*})=0$. This is the condition of macroscopic reversibility \cite{MFT} that corresponds microscopically to the inhomogeneous reversible product measure discussed at the end of section \ref{unno}. In this case the quasi-potential $V_{E^*}$ is local and can be computed both microscopically like in section \ref{trc} that macroscopically using \eqref{HJ}. We obtain that $V_{E^*}(\rho)=\mathcal V_{\lambda\left[\bar\rho_{E^*}\right]}(\rho)$ and $\lambda\big[\bar\rho{_{E^*}}(x)\big]$ linearly interpolates $\rho_-$ and $\rho_+$ when $x\in[0,1]$.

\subsection{The case $E<E^*$}

For some computations it is more convenient to use the variable $\psi=\lambda[\phi]$.  Instead of discuss the general case we consider  the three prototype models.
Recall that in the case of the KMP model this change of variables corresponds to $\psi=-\frac 1\phi$.
We discuss before the KMP case showing then how to modify the computations to cover also the other cases.
We consider the functional \eqref{effeciu} with a fixed determination of the signs of the moduli in \eqref{arrr}. This is enough to identify the correct solution. We write  the functional in terms of the variable $\psi$. The choice of the sign of the two logarithmic terms has to be $+$ since otherwise the function $\psi$ cannot satisfy the boundary conditions. Consequently we have to restrict the domain of definition of $\mathcal G_E$. We add also a suitable constant to fix the normalization. In terms of $\psi$ the functional becomes
\begin{eqnarray}\label{eg}
\mathcal G_E(\rho,\psi)&=&\int_\Lambda\left[\left(\frac{\nabla\psi}{E}-1\right)\log\left(\nabla\psi-E\right)
-\frac{\nabla\psi}{E}\log \left(\nabla\psi\right)\right]dx\nonumber \\
& +&\int_\Lambda\left[-\rho\psi +\log \left(-\frac{\psi}{\rho}\right)-1\right]dx+K_E
\end{eqnarray}
where the constant $K_E$ is
\begin{equation}\label{Kekmp}
K_E=\log\left(-J_E\right)+\frac 1E\int_{\rho_-}^{\rho_+}\frac{d \rho}{\sigma(\rho)}\log\left(1-\frac{\sigma(\rho)E}{J_E}\right)\,.
\end{equation}
The functions $\psi$ that we are considering belong to
\begin{equation}\label{domps}
\mathcal F_E:=\left\{\psi\in C^1(\Lambda)\,:\,\nabla\psi\geq \max\{E,0\}\,,\, \psi(0)=-\frac{1}{\rho_-}\,, \psi(1)=-\frac{1}{\rho_+}\right\}\,.
\end{equation}
For a $\psi \in \mathcal F_E$ the functional \eqref{eg} is well defined. Formula \eqref{eg} is not well defined in the special case $E=0$. This case corresponds to the symmetric KMP process and has been already discussed in \cite{BGL}. It is possible to obtain the corresponding functional for this special case as a limit of \eqref{eg} when $E\to 0$.
The constant $K_E$ has been fixed in such a way that $\inf_{\rho,\psi\in \mathcal F_E}\mathcal G_E(\rho,\psi)=0$. We discuss later in the general framework this point.

For any fixed $\rho$ the functional $\mathcal G(\rho,\cdot)$ is neither concave nor convex. Its critical points are determined by the Euler-Lagrange equation
\begin{equation}\label{elkmmm}
\frac{\Delta \psi}{\nabla\psi(E-\nabla\psi)}+\frac{1}{\psi}=\rho\,.
\end{equation}
We define the functional
\begin{equation}\label{esed1}
S_E(\rho)=\inf_{\psi\in \mathcal F_E} \mathcal G_E(\rho, \psi)\,.
\end{equation}
We can identify the quasi-potential $W_E=V_E$ with the infimum \eqref{esed1}, i.e. $V_E=W_E=S_E$. This is based on the interpretation of $\mathcal G_E$ as the pre-potential in an Hamiltonian framework obtained interpreting \eqref{drf} as a Lagrangian action \cite{Lag,CPAM,MFT}.
The pre-potential is defined on the unstable manifold for the Hamiltonian flow relative to a suitable equilibrium point  associated to the stationary solution $\bar\rho_E$. The value $\mathcal G_E(\rho,\psi)$ coincides with the value of the pre-potential when the pair $(\rho,\psi)$ belongs to the unstable manifold. Since the unstable manifold can be characterized by the stationary condition \eqref{elkmmm} we can then consider simply the infimum in \eqref{esed1} since all the critical points are belonging to the unstable manifold. For these values of the field $E$ it is not possible to show that there is uniqueness in the minimizer in \eqref{qp} and equivalently in \eqref{esed1}. In correspondence the unstable manifold is not a graph and there is the possibility to have Lagrangian phase transitions. We are not going to discuss the details of these arguments and we refer to \cite{Lag,CPAM} for the analogous computation in the case of the exclusion.

For the KMPd model we use again the change of variables $\psi=\lambda[\phi]=\log\frac{\phi}{\phi+1}$ and the corresponding functional $\mathcal G_E$ has a form very similar to \eqref{eg}. In particular we give a general form that works for all the three models that we are considering and that depends on the density of free energy and the transport coefficients. In particular  \eqref{eg} can be obtained as a special case. The general form is constituted by three terms. The first one depends only on $\nabla \psi$ and coincides with the first term in \eqref{eg}. The second one can be written in general as
\begin{equation}\label{egdd}
\int_\Lambda\left[f(\rho)-f\left(\rho[\psi]\right)-\psi\big(\rho-\rho[\psi]\big)-\log\sigma\big(\rho[\psi]\big)\right] \,dx\,.
\end{equation}
The third term is the constant \eqref{Kekmp}.
When $\rho$ is fixed the general form of the Euler Lagrange equation for $\mathcal G_E(\rho,\cdot)$ can be written in the general form
\begin{equation}\label{eld}
\frac{\Delta\psi}{\nabla\psi(E-\nabla\psi)}+\rho[\psi]-\sigma'\big(\rho[\psi]\big)=\rho\,.
\end{equation}
These general expressions work also for the KMPx model. The change of variable is again $\psi=\lambda[\phi]=\arctan \phi$.

The functional space for the $\psi$ is like \eqref{domps} with just a difference in the boundary values. In particular for KMPd we have $\psi(0)=\log\frac{\rho_-}{1+\rho_-}$ and $\psi(1)=\log\frac{\rho_+}{1+\rho_+}$ while for KMPd we have $\psi(0)=\arctan\rho_-$ and $\psi(1)=\arctan\rho_+$.

\smallskip

To compute the global infimum of $\mathcal G_E$, that is relevant for the determination of the normalizing constant, it is convenient to minimize before in $\rho$ keeping fixed $\psi$. The stationary condition that corresponds to a minimum is $\lambda[\rho]=\psi$ and in correspondence the term \eqref{egdd} reduces to
$-\int_\Lambda\log\sigma\big(\rho[\psi]\big) \,dx$.
We minimize now over $\psi$ and obtain the stationary condition
\begin{equation}\label{psdc}
\frac{\Delta\psi}{\nabla\psi(E-\nabla\psi)}=\sigma'\big(\rho[\psi]\big)\,.
\end{equation}
Using the change of variable $\rho=\rho[\psi]$ equation \eqref{psdc} becomes the stationary equation in \eqref{steq} that has an unique solution. We obtain then $\inf_{\rho,\psi\in\mathcal F_E}\mathcal G_E(\rho,\psi)=\mathcal G_E\left(\bar\rho_E,\lambda\big[\bar\rho_E\big]\right)$ and imposing that this value is zero we get the general formula \eqref{Kekmp}.

\subsection{The case $E>E^*$}
Also in this case we use the variable $\psi=\lambda[\phi]$. We consider the functional \eqref{effeciu} in terms of this new variable. In this case the sign of the modulus in the second logarithmic term in \eqref{arr} is still $+$ while the first one has to be fixed as $-$. This is because the values of the field are different and this is the choice that allows to satisfy the boundary conditions for $\psi$. Consequently we have to restrict the functions considered. In the case of the KMP model we have
\begin{eqnarray}\label{zanna}
\mathcal G_E(\rho,\psi)&=&\int_\Lambda\left[\left(\frac{\nabla\psi}{E}-1\right)\log(E-\nabla\psi)
-\frac{\nabla\psi}{E}\log \nabla\psi\right]dx\nonumber \\
& +&\int_\Lambda\left[-\rho\psi +\log \left(-\frac{\psi}{\rho}\right)-1\right]dx+K_E\,,
\end{eqnarray}
where the constant $K_E$ is
\begin{equation}\label{Kekmp2}
K_E=\log(J_E)+\frac 1E\int_{\rho_-}^{\rho_+}\frac{d \rho}{\sigma(\rho)}\log\left(\frac{\sigma(\rho)E}{J_E}-1\right)\,.
\end{equation}
The function $\psi$ is a function belonging to the set
\begin{equation}\label{doveleo}
\mathcal F_E:=\left\{\psi\in C^1(\Lambda)\,:\, 0\leq\nabla\psi\leq E\,,\, \psi(0)=-\frac{1}{\rho_-}\,,\psi(1)=-\frac{1}{\rho_+}\right\}\,.
\end{equation}
We have that the function $\left(\frac{\alpha}{E}-1\right)\log(E-\alpha)
-\frac{\alpha}{E}\log \alpha$ is concave when $\alpha \in [0,E]$ and also the function $\log (-\alpha)$
defined on the negative real line. The concavity of the functions implies the concavity of the functional $\mathcal G_E(\rho,\cdot)$ for any fixed $\rho$. Like in \cite{BGLa} there exists in $\mathcal F_E$ an unique critical point of $\mathcal G_E(\rho,\cdot)$ that is then a maximum. This is obtained as the unique solution in $\mathcal F_E$ to the Euler-Lagrange equation
\begin{equation}\label{ELK}
\frac{\Delta\psi}{\nabla\psi(\nabla\psi-E)}+\frac{1}{\psi}=\rho\,.
\end{equation}
We define the functional
\begin{equation}\label{esed}
S_E(\rho)=\sup_{\psi\in \mathcal F_E} \mathcal G_E(\rho, \psi)=\mathcal G_E(\rho,\psi^\rho)
\end{equation}
where $\psi^\rho$ is the maximizer solving \eqref{ELK}. Again we have $S_E=V_E=W_E$.
In this case the uniqueness of the solution of \eqref{ELK} is related to the fact that there is an unique critical point
for the variational problem related to the quasi-potential \eqref{qp}. The minimizer can be better identified in terms of the variable $\phi$. As discussed in \cite{BGLa,MFT} the minimizer for the computation of $V_E(\rho^*)$, where $\rho^*$ is a generic density profile, is characterized
by the time reversal of the $\rho$ solving the following differential problem
\begin{equation}\label{tmc}
\left\{
\begin{array}{ll}
\partial_t \rho= \Delta \rho-\nabla\cdot\left(\rho^2\left(E-\frac{\nabla\phi}{\phi^2}\right)\right)\,, & (x,t)\in \Lambda\times (0,+\infty)\,,\\
\frac{\phi^2\Delta\phi-2\phi\left(\nabla\phi\right)^2}{\nabla\phi\left(\nabla\phi-\phi^2E\right)}-\phi=\rho\,, & (x,t)\in \Lambda\times (0,+\infty)\,,\\
\rho(x,0)=\rho^*(x)\,, & \\
\rho(0,t)=\phi(0,t)=\rho_-\,,\rho(1,t)=\phi(1,t)=\rho_+\,. &
\end{array}
\right.
\end{equation}
The second equation in \eqref{tmc} is \eqref{ELK} written in terms of the variable $\phi$. The first equation in \eqref{tmc} is the hydrodynamic equation associated to a weakly asymmetric model having a space and time dependent external field $F=E-\frac{\nabla\phi}{\phi^2}$. The value of this external field depends on the density $\rho$ and has to be computed solving the second equation in \eqref{tmc}. By the general theory \cite{MFT} the $\rho$ solving \eqref{tmc} is the solution of the hydrodynamic equation associated to the time reversed process. It is possible to check that the differential problem \eqref{tmc} is equivalent to the simpler differential problem
\begin{equation}\label{tmcs}
\left\{
\begin{array}{ll}
\partial_t \phi=\Delta\phi-\nabla(\phi^2E)\,, & (x,t)\in \Lambda\times (0,+\infty)\,,\\
\frac{\phi^2\Delta\phi-2\phi\left(\nabla\phi\right)^2}{\nabla\phi\left(\nabla\phi-\phi^2E\right)}-\phi=\rho\,, & (x,t)\in \Lambda\times (0,+\infty)\,,\\
\rho(x,0)=\rho^*(x)\,, & \\
\rho(0,t)=\phi(0,t)=\rho_-\,,\rho(1,t)=\phi(1,t)=\rho_+\,, &
\end{array}
\right.
\end{equation}
where $\phi$ solves an autonomous equation that indeed coincide with the original hydrodynamic equation. The minimizer for the computation of the quasi-potential is obtained by the time reversal of the solution $\rho$ of \eqref{tmcs}. Also in this case $\mathcal G_E$ can be interpreted as the pre-potential in an Hamiltonian framework but in this case for any $\rho$ there is an unique $(\rho,\psi)$ belonging to the unstable manifold. We have then not to minimize over the different points of the unstable manifold. It turn out that we have instead to maximize because the unstable manifold is exactly characterized by the stationary condition \eqref{ELK} and the critical point corresponds to a maximum by concavity \cite{BGLa}.

Also in this case we can obtain similar expressions of $\mathcal G_E$ for the models KMPd and KMPx and write a general expression for it.
Like in the previous case we have that $\mathcal G_E$ is composed by the sum of 3 terms. The first one depending only on $\nabla \psi$ coincides with the first term in \eqref{zanna}. The second term has a general form that coincides with \eqref{egdd}. The additive constant has the form \eqref{Kekmp2}. Also for these models the functional $\mathcal G_E(\rho,\cdot)$ is concave.

\section{Totally asymmetric limit}

In this section we study the asymptotic limit of the quasi-potential when the external field is large. This is done, like in \cite{Lag,CPAM,BGLa}, studying before the limit of $\mathcal G_E$ and then solving a corresponding variational problem.

\subsection{The case $E\to -\infty$}

We study first the limit when $E\to -\infty$ of the auxiliary functional $\mathcal G_E$. Since we are interested in the limiting value we can assume that $E<0$ and it is convenient to add and subtract the term $\log (-E)$ in \eqref{eg}. We add this factor to the first term of \eqref{eg} that becomes $\int_\Lambda s\left(-\frac{\nabla\psi}{E}\right)dx$ with
\begin{equation}\label{vitam}
s(\alpha):=\alpha\log\alpha-(1+\alpha)\log(1+\alpha)\,.
\end{equation}
Since $\lim_{\alpha\downarrow 0}s(\alpha)=0$ this term is converging to zero in the limit of large and negative field. The factor that we subtract is inserted in the additive constant $K_E$ that becomes
\begin{equation}\label{riu1}
\log\frac{J_E}{E}+\frac 1E\int_{\rho_-}^{\rho_+}\frac{d \rho}{\sigma(\rho)}\log\left(1-\frac{\sigma(\rho)E}{J_E}\right)\,.
\end{equation}
The asymptotic behavior of \eqref{riu1} can be easily understood since several terms depend just on the ratio $\frac{J_E}{E}$ whose behavior in the limit for large fields is given by \eqref{bgn}. In particular in this case the second term in \eqref{riu1} converges to zero while the first one converges to $2\log \rho_+$.

The second term in \eqref{eg} does not depend on $E$ and remains identical in the limit. We obtained then that
$\mathcal G_{-}:=\lim_{E\to -\infty}\mathcal G_E$ is given by
\begin{equation}\label{defgmeno1}
\mathcal G_-(\rho,\psi)=\int_0^1\left[-\rho\psi +\log \left(-\frac{\psi}{\rho}\right)-1\right]dx+2\log\rho_+\,.
\end{equation}
The function $\psi$ in \eqref{defgmeno1} has to belong to the set
\begin{equation}\label{stpiz}
\mathcal F_-:=\left\{\psi\in C^1(\Lambda)\,:\, \nabla\psi\geq 0\,, \psi(0)=-\frac{1}{\rho_-},\psi(1)=-\frac{1}{\rho_+},\right\}\,.
\end{equation}
We can obtain (see \cite{CPAM,BGLa}) $V_-(\rho):=\lim_{E\to-\infty}V_E(\rho)$ as
\begin{equation}
V_-(\rho)=S_-(\rho):=\inf_{\psi\in \mathcal F_-}\mathcal G_-(\rho,\psi)\,.
\end{equation}
Since the function
$\psi\to-\rho\psi +\log \left(-\frac{\psi}{\rho}\right)$ is decreasing when
$\psi \in\left[\frac{-1}{\rho_-},\frac{-1}{\rho_+}\right]$, the minimum  of the integrand in \eqref{defgmeno1} over the $\psi$ for a fixed $\rho$ is obtained for  $\psi=-\frac{1}{\rho_+}$. This means that we can construct a minimizing sequence in $\mathcal F_-$ approximating a function that takes the value $-\frac{1}{\rho_-}$ at $0$ and then immediately jumps to the value $-\frac{1}{\rho_+}$. We then obtain
\begin{equation}\label{maleo}
V_-(\rho)=\inf_{\psi\in \mathcal F_-} \mathcal G(\rho,\psi)=\int_\Lambda \left[\frac{\rho}{\rho_+}-\log\frac{\rho}{\rho_+}-1\right]dx
\end{equation}
that is the large deviation rate function for masses distributed according to a product of exponentials distributions of parameter $\frac{1}{\rho_+}$. This is  the large deviations rate functional for the invariant measure of the version 1 of the totally asymmetric KMP dynamics discussed in Section \ref{tkmp1dis} if we invert the direction of the asymmetry there.

\smallskip

For the KMPd model the asymptotic behavior is quite similar. The limiting value of $\frac{J_E}{E}$ when $E\to -\infty$ coincides always with $\rho_+(1+\rho_+)$.
 This implies that the limiting value of the constant term \eqref{riu1} is equal to $\log(\rho_+(1+\rho_+))$. The term depending only on $\nabla \psi$ is the same as in the KMP case and it is converging to zero. The term \eqref{egdd} does not depend on $E$ and is preserved in the limit. We obtain in the limit
\begin{equation}\label{limgd-}
\mathcal G_-(\rho,\psi)=\int_0^1\left[f(\rho)+\log(1-e^\psi)-\psi(\rho+1)\right]dx +\log(\rho_+(1+\rho_+))\,,
\end{equation}
and $\psi$ belongs to
\begin{equation}\label{stpizd}
\mathcal F_-:=\left\{\psi\in C^1(\Lambda)\,:\, \nabla\psi\geq 0\,, \psi(0)=\log\frac{\rho_-}{1+\rho_-},\psi(1)=\log\frac{\rho_+}{1+\rho_+},\right\}\,.
\end{equation}
As before in the interval $\left[\log\frac{\rho_-}{1+\rho_-},\log\frac{\rho_+}{1+\rho_+}\right]$ the function $\psi\to\log(1-e^\psi)-\psi(\rho+1)$ is decreasing  and then a minimizing sequence in $\mathcal F_-$ is converging
to a function that is equal to $\log\frac{\rho_-}{1+\rho_-}$ at $0$ and then is constantly equal to $\log\frac{\rho_+}{1+\rho_+}$. We obtain \cite{CPAM,BGLa} that $V_-=\lim_{E\to -\infty}V_E$ is obtained by
\begin{equation}\label{evaid}
V_-(\rho)=S_-(\rho):=\inf_{\psi\in \mathcal F_-} \mathcal G_-(\rho,\psi)=\int_\Lambda\left[\rho\log\frac{\rho}{\rho_+}-(1+\rho)\log\frac{1+\rho}{1+\rho_+}\right]
\end{equation}
that is the large deviation rate functional for a product measure with marginals given by geometric distributions of
parameter $\frac{1}{1+\rho_+}$.

\smallskip

The limiting behavior of the KMPx model is instead quite different and exhibits a behavior similar to the exclusion process.
Recalling \eqref{bgn}, for the KMPx model the limiting value of $\frac{J_E}{E}$ when $E\to -\infty$  is given by $\max\{1+\rho_-^2,1+\rho_+^2\}$. Let us call
\begin{equation}\label{m}
\bar\rho:=\left\{
\begin{array}{ll}
\rho_- & \textrm{if}\ |\rho_-|\geq |\rho_+|\,, \\
\rho_+ & \textrm{if}\ |\rho_-|< |\rho_+|\,.
\end{array}
\right.
\end{equation}
The limiting functional $\mathcal G_-$ in this case becomes
\begin{equation}\label{arrd}
\mathcal G_-(\rho,\psi)=\int_0^1\left[\rho\arctan\rho-\frac{\log(1+\rho^2)}{2}-\rho\psi-\frac{\log(1+(\tan\psi)^2)}{2}\right]dx
+\log(1+\bar\rho^2)\,,
\end{equation}
and the function $\psi$ has to belong to the set
\begin{equation}\label{stpizx}
\mathcal F_-:=\left\{\psi\in C^1(\Lambda)\,:\, \nabla\psi\geq 0\,, \psi(0)=\arctan(\rho_-),\psi(1)=\arctan(\rho_+)\right\}\,.
\end{equation}
Since the function $\psi\to-\log(1+(\tan\psi)^2)$ is concave
the functional $\mathcal G_-(\rho,\cdot)$ is also concave.
To compute $V_-(\rho)=S_-(\rho)$ we need to minimize  $\mathcal G_-(\rho,\cdot)$ over the convex set \eqref{stpizx}. The infimum is
then realized on the extremal points of a suitable closure of that convex set. The extremal elements are the function of the form
\begin{equation}\label{step}
\psi^y(x)=\arctan(\rho_-)\chi_{[0,y)}(x)+\arctan(\rho_+)\chi_{[y,1]}(x)\,,\qquad y\in(0,1]\,.
\end{equation}
This means piecewise constant functions jumping from $\arctan(\rho_-)$ to $\arctan(\rho_+)$ at the single point $y$. Let us define the functional
\begin{eqnarray}\label{dinotte}
& &\tilde{\mathcal G}_-(\rho,y):=\mathcal G_-(\rho,\psi^y)=\int_0^1\left[\rho\arctan\rho
-\frac 12\log(1+\rho^2)\right]dx\nonumber \\
& &-\arctan(\rho_-)\int_0^y\rho(x)\, dx-\arctan(\rho_+)\int_y^1\rho(x)\, dx\nonumber \\
& &-\frac{y}{2}\log(1+\rho_-^2)-\frac{1-y}{2}\log(1+\rho_+^2)+\log(1+\bar\rho^2)\,.
\end{eqnarray}
This is the functional $\mathcal G_-$ computed in correspondence of the extremal elements.
The infimum of $\mathcal G_-(\rho,\cdot)$ coincides with the infimum of the function of one real variable
$\tilde{\mathcal G}_-(\rho,\cdot)$ on the interval $[0,1]$. Consequently we have
$$
V_-(\rho)=S_-(\rho)=\inf_{y\in(0,1]}\tilde{\mathcal G}_-(\rho,y)\,.
$$
The stationary condition for $\tilde{\mathcal G}_-(\rho,\cdot)$ is
\begin{equation}\label{stgt}
\rho(y)\left[\arctan\rho_+-\arctan\rho_-\right]=\frac 12 \log\frac{1+\rho_-^2}{1+\rho_+^2}\,.
\end{equation}
The second derivative establishing if a critical point is a local minimum or a local maximum
is given by $\nabla\rho(y)\left[\arctan\rho_+-\arctan\rho_-\right]$. To find the global minimum we have to consider also the values of the functions on the two extrema of the interval. Like for the exclusion process it is possible to have that the minimum
is obtained in more than one single point $y$ and this phenomenon is suggesting the possibility of the existence of dynamic phase transitions \cite{Lag} for finite and large enough negative fields.

\subsection{The case $E\to +\infty$}

We study first the limit when $E\to +\infty$ of the auxiliary functionals $\mathcal G_E$ \cite{CPAM,BGLa}. We start with the KMP model. As before it is convenient to add and subtract the term $\log E$. We add this factor to the first term of \eqref{zanna} that becomes $\int_\Lambda \tilde s\left(\frac{\psi_x}{E}\right)dx$ with
\begin{equation}
\tilde s(\alpha)=-\alpha\log\alpha-(1-\alpha)\log(1-\alpha)
\end{equation}
Since $\lim_{\alpha\downarrow 0}\tilde s(\alpha)=0$ this term is converging to zero in the limit of large and positive field. The factor that we subtract is inserted in the additive constant $K_E$ and we obtain
\begin{equation}\label{riu}
\log\frac{J_E}{E}+\frac 1E\int_{\rho_-}^{\rho_+}\frac{d \rho}{\sigma(\rho)}\log\left(\frac{\sigma(\rho)E}{J_E}-1\right)\,.
\end{equation}
The asymptotic behavior of \eqref{riu} can be easily understood since several terms depend just on the ratio $\frac{J_E}{E}$ whose behavior in the limit for large fields is given by \eqref{bgn}. In particular in the KMP case the second term in \eqref{riu} converges to zero while the first one converges to $2\log \rho_-$.

The second term in \eqref{zanna} does not depend on $E$ and remains identically in the limit. We deduce that
$\lim_{E\to +\infty}\mathcal G_E=\mathcal G_{+}$ defined as
\begin{equation}\label{defgmeno}
\mathcal G_+(\rho,\psi)=\int_0^1\left[-\rho\psi +\log \left(-\frac{\psi}{\rho}\right)-1\right]dx+2\log\rho_-\,.
\end{equation}
Since we are considering the limit of a large positive field the bound from above on the derivative in \eqref{doveleo} disappears
and the function $\psi$ in \eqref{defgmeno} belongs to the set
\begin{equation}\label{dovelea3}
\mathcal F_+:=\left\{\psi\in C^1(\Lambda)\,:\, \nabla\psi\geq 0\,, \psi(0)=-\frac{1}{\rho_-}\,,\psi(1)=-\frac{1}{\rho_+}\right\}\,.
\end{equation}
Since the function
$\psi\to-\rho\psi +\log \left(-\frac{\psi}{\rho}\right)$ is decreasing when
$\psi \in\left[-\frac{1}{\rho_-},-\frac{1}{\rho_+}\right]$ we can compute the supremum  of \eqref{defgmeno} over the $\psi$ for a fixed $\rho$ obtaining a minimizing sequence converging to a function that assumes the value $-\frac{1}{\rho_-}$ on all the interval except the point $1$ where it assumes the value $-\frac{1}{\rho_+}$. We then obtain
\begin{equation}\label{maleo+}
V_+(\rho)=S_+(\rho):=\sup_{\psi\in \mathcal F_+} \mathcal G_+(\rho,\psi)=\int_0^1 \left[\frac{\rho}{\rho_-}-\log\frac{\rho}{\rho_-}-1\right]dx
\end{equation}
that is the large deviation rate function for masses distributed according to a product of exponentials of parameter $\frac{1}{\rho_-}$. This is exactly the large deviations rate functional for the invariant measure of the version 1 of the totally asymmetric KMP dynamics discussed in Section \ref{tkmp1dis}.

\smallskip

For the KMPd model the limiting value of $\frac{J_E}{E}$ when $E\to +\infty$ coincides always with $\rho_-(1+\rho_-)$.
The limiting behavior for this model is very similar to the KMP. In particular the limiting value of the constant term \eqref{riu} is equal to $\log(\rho_-(1+\rho_-))$. The term depending only on $\nabla \psi$ is the same as the KMP case and it is converging to zero. The term \eqref{egdd} does not depend on $E$ and is preserved in the limit. Writing explicitly the terms we obtain in the limit
\begin{equation}\label{limgd}
\mathcal G_+(\rho,\psi)=\int_0^1\left[f(\rho)+\log(1-e^\psi)-\psi(\rho+1)\right]dx +\log(\rho_-(1+\rho_-))
\end{equation}
The function $\psi$ in \eqref{limgd} belongs to the set
\begin{equation}\label{dovelea2}
\mathcal F_+:=\left\{\psi\in C^1(\Lambda)\,:\, \nabla\psi\geq 0\,, \psi(0)=\log\frac{\rho_-}{1+\rho_-}\,,\psi(1)=\log\frac{\rho_+}{1+\rho_+}\right\}\,.
\end{equation}
Since in the interval $\left[\log\frac{\rho_-}{1+\rho_-},\log\frac{\rho_+}{1+\rho_+}\right]$ the function $\psi\to\log(1-e^\psi)-\psi(\rho+1)$ is decreasing,
the supremum to compute $V_+=S_+$ is realized on a function $\psi$ that is constantly equal to $\log\frac{\rho_-}{1+\rho_-}$. We obtain
\begin{equation}\label{evaid-}
V_+(\rho)=S_+(\rho):=\sup_{\psi\in \mathcal F_+} \mathcal G_+(\rho,\psi)=\int_0^1\left[\rho\log\frac{\rho}{\rho_-}-(1+\rho)\log\frac{1+\rho}{1+\rho_-}\right]
\end{equation}
that is the large deviation rate functional for a product measure with marginals given by geometric distributions of
parameter $\frac{1}{1+\rho_-}$.

\smallskip

For the KMPx model the limiting value of $\frac{J_E}{E}$ when $E\to +\infty$ has 3 different possible values.
When $0\leq\rho_-\leq\rho_+$ then the limiting value is $1+\rho_-^2$. When $\rho_-\leq0\leq\rho_+$ then the limiting value is $1$. When $\rho_-\leq \rho_+\leq 0$ then the limiting value is $1+\rho_+^2$. Let us define
\begin{equation}
\bar\rho:=\left\{
\begin{array}{ll}
\rho_- & \textrm{if}\ 0\leq\rho_-\leq\rho_+\,, \\
0 & \textrm{if}\  \rho_-\leq0\leq\rho_+\,,\\
\rho_+ & \textrm{if}\ \rho_-\leq \rho_+\leq 0\,.
\end{array}
\right.
\end{equation}
With this definition we have  $\frac{J_E}{E}\to\sigma\left(\bar\rho\right)$.
The limiting functional $\mathcal G_+$ in this case becomes
\begin{equation}\label{arr}
\mathcal G_+(\rho,\psi)=\int_0^1\left[\rho\arctan\rho-\frac{\log(1+\rho^2)}{2}-\rho\psi-
\frac{\log(1+(\tan\psi)^2)}{2}\right]dx+\log(1+(\bar\rho)^2)\,.
\end{equation}
and the function $\psi$ belongs to
\begin{equation}\label{corri}
\mathcal F_+:=\left\{\psi\in C^1(\Lambda)\,:\, \nabla\psi\geq 0\,, \psi(0)=\arctan\rho_-\,,\psi(1)=\arctan\rho_+\right\}\,.
\end{equation}
We have $V_+(\rho)=S_+(\rho)=\sup_{\psi\in \mathcal F_+} \mathcal G_+(\rho,\psi)$ and the maximizer $\psi_M^\rho$ such that $V_+(\rho)=\mathcal G_+\left(\rho,\psi_M^\rho\right)$ can be described as follows. Let $H(x)=-\int_0^x\rho(y)\,dy$ and $G=co(H)$ its convex envelope. We define
\begin{equation}\label{lafi}
\phi_M^\rho(x)=\left\{
\begin{array}{ll}
\rho_- & \textrm{if}\ \nabla G(x)\leq\rho_- \\
\rho_+ & \textrm{if}\ \nabla G(x)\geq\rho_+ \\
\nabla G(x) & \textrm{otherwise}\,.
\end{array}
\right.
\end{equation}
We have that the maximizer is $\psi_M^\rho=\arctan\phi_M^\rho$. To prove this statement it is convenient to write back \eqref{arr} in terms of the variable $\phi$ related to $\psi$ by $\phi=\tan\psi$. We need to prove that $\phi_M^\rho$
is the minimizer of
\begin{equation}\label{irafunesta}
\inf_{\phi\in \tilde{\mathcal F}_+} \int_0^1 \left[\rho \arctan\phi+\frac 12\log \left(1+\phi^2\right)\right]dx=
\inf_{\phi\in \tilde{\mathcal F}_+} \mathcal B(\rho,\phi)\,,
\end{equation}
where the last equality defines the functional $\mathcal B$ and
\begin{equation}\label{nes}
\tilde{\mathcal F}_+:=\left\{\phi\in C^1(\Lambda)\,:\, \nabla\phi\geq 0\,, \phi(0)=\rho_-\,,\phi(1)=\rho_+\right\}\,.
\end{equation}
To identify the minimizer $\phi_M^\rho$ in \eqref{irafunesta} we can adapt the argument in \cite{DLS}. Since $H(0)=G(0)$ and $H(1)=G(1)$ with an integration by parts we obtain
\begin{equation}\label{partii}
\int_0^1\left(\rho+\nabla G\right)\arctan(\phi)\, dx=\int_0^1\left(H-G\right)\nabla \arctan(\phi)\, dx\geq 0\,.
\end{equation}
The last inequality follows by the fact that $\nabla \phi\geq 0 $, the function $\arctan$ is increasing and by definition $G\leq H$. We have that the second term of \eqref{partii} is the integral of the product of two non negative terms and consequently it is non negative. Inequality \eqref{partii} can be written as $\mathcal B(\rho,\phi)\geq \mathcal B(-\nabla G,\phi)$.

The derivative with respect to $\phi$ of the function $\alpha \arctan\phi+\frac 12\log \left(1+\phi^2\right)$ is given by $\frac{\alpha+\phi}{1+\phi^2}$ and is negative for $\phi<-\alpha$ and positive for $\phi>-\alpha$. Since the function $\phi(x)$ takes values only on the interval $[\rho_-,\rho_+]$ we have that
$$
\inf_{\phi\in [\rho_-,\rho_+]}\left\{\alpha \arctan\phi+\frac 12\log \left(1+\phi^2\right)\right\}
$$
is obtained at $\rho_-$ when $\alpha\geq -\rho_-$, it is obtained at $\rho_+$ when $\alpha\leq -\rho_+$ and it is obtained at $-\alpha$ when $\alpha\in [-\rho_+,-\rho_-]$. This fact implies that we have the inequality $\mathcal B(-\nabla G,\phi)\geq \mathcal B(-\nabla G,\phi_M^\rho)$. The last fact that remains to prove is that it holds the equality $\mathcal B(-\nabla G,\phi_M^\rho)= \mathcal B(\rho,\phi_M^\rho)$. This follows by the following argument. We split the interval $[0,1]$ on intervals where either $-\nabla G=\rho$ or $\nabla G$ is constant. On an interval where $-\nabla G=\rho$ we have obviously the same contribution. On an interval $[a,b]$ where $\nabla G$ is constant correspondingly also $\phi_M^\rho$ is constant. Moreover the constant value of $\nabla G$ coincides with $\frac{\int_a^b\rho(y)\,dy}{a-b}$. Since $\phi_M^\rho$ is constant on the interval the contribution coming from intervals of this type still coincide and we get the equality.
Summarizing we have the following chain
\begin{equation}\label{chaina}
\mathcal B(\rho,\phi)\geq \mathcal B(-\nabla G,\phi)\geq \mathcal B(-\nabla G,\phi_M^\rho)= \mathcal B(\rho,\phi_M^\rho)
\end{equation}
that implies that $\phi_M^\rho$ is the minimizer in \eqref{irafunesta}. Since $\psi=\arctan \phi$ we obtain that the maximizer to compute $V_+(\rho)=S_+(\rho)$ is given by $\psi_M^\rho=\arctan\phi_M^\rho$.

\subsection*{Acknowledgements}
D.G. acknowledges L. Bertini, A. De Sole, G. Jona Lasinio and C. Landim for a long collaboration on the general subject. We especially thank C. Bernardin and L. Bertini for very useful discussions on this specific problem and for informing us of their related computations \cite{BP,bpc}.

\end{document}